\documentstyle[twocolumn,aps,prc,epsf]{revtex}
\begin{document}
\draft
\title{Analytic proof that the Quark Model complies with the PCAC theorems}
\author{Pedro Bicudo}
\address{
Departamento de F\'{\i}sica, and CFIF, Instituto Superior T\'ecnico,
Av. Rovisco Pais, 1049-001 Lisboa, Portugal
}
\maketitle
\begin{abstract} 
The Weinberg theorem, the Adler self consistency zero, the 
Goldberger and Treiman relation and the Gell-Mann Oakes and 
Renner relation are proved analytically in full detail  
for Quark Models. 
These proofs are independent of the particular quark-quark interaction,
and they are displayed with Feynman diagrams in a compact notation.
I assume the ladder truncation, which is natural in the Quark Model,
and also detail the diagrams that must be included in each relation.
Off mass shell and finite size effects are included in the quark-antiquark 
pion Bethe Salpeter vertices. 
The axial and vector Ward identities, for the quark propagator and 
for the ladder, exactly cancel any model dependence.
\end{abstract}
\pacs{ }
%

\section{Introduction} 

\par
The pion was introduced by Yukawa in 1931,
to account for the strong nucleon-nucleon attraction which
binds the nucleus. Yukawa was inspired by the Coulomb attraction
in atomic physics which is due to the photon
exchange interaction. The pion was indeed experimentally 
discovered, it is a pseudoscalar and an isovector. The pion mass
$M_{\pi^\pm}=140\, MeV$ and $M_{\pi^0}=135\, MeV$ determines
the range of the nucleon-nucleon attraction and confirms 
the prediction of Yukawa. 
The analogy with photon physics went quite far. 
The $U(1)$ gauge symmetry 
is a crucial property of Quantum Electrodynamics.
In hadronic physics there is also a symmetry, chiral symmetry, which
is a spontaneously broken global symmetry. 
In the chiral limit (limit of exact chiral symmetry) the pion would
play the role of the massless Goldstone boson.
The pion mass is finite but it is indeed much smaller than 
the mass scale of hadronic physics which is of the GeV order.
The expansion in the pion mass, together with the 
techniques of current algebra led to beautifully correct 
theorems, the  PCAC (Partially Conserved Axial Current) theorems. 
Similar to the vector Ward identities in gauge symmetry,
the axial Ward identities constitute a powerful tool of 
chiral symmetry. 
An important parameter of PCAC is $f_\pi$
which relates the pion vertex with the axial vertex.
$f_\pi=93 MeV$ is measured in the electroweak pion decay,
and it is also known as the pion decay constant.

\par
Other hadrons, including hundreds of resonances
were also found subsequently. The large number of
hadrons, and Deep Inelastic Scattering, led to the discovery
of quarks and to QCD (Quantum Chromodynamics) which is the presently
accepted theory of strong interactions. QCD has not been solved yet, but it
inspired the invention of the Quark Model \cite{Rujula}
in order to describe the bound states of quarks, which fall mainly in the 
classes of mesons, like the pion, and baryons, like the proton. 
The Quark Model also uses confining potentials which are determined in lattice QCD. 
The success of the Quark Model relies on its ability to reproduce 
the whole spectrum of hadronic resonances, with microscopic interacting quarks. 
Moreover the Quark Model is competent to explain microscopically the
strong hadron-hadron elastic interactions
\cite{Ribeiro}.
For recent coupled channel studies see
\cite{Gastao,pi-pi,Goncalo}. 

\par
However the Quark Model suffered from the onset 
to accommodate the low pion mass. The mass scale of hadronic
physics is of the $GeV$ order.
The Quark Model needed a large number of 
parameters in order to fit the low pion mass, 
and to address pion creation and annihilation 
in hadronic decays. It is clear that a light pion is natural
in chiral physics, while it is odd in constituent models.
With the aim to cure the important problems
of the pion mass \cite{Pene,Adler2}, of the pion coupling \cite{Bicudo},
and of the vacuum condensate \cite{Bicudo}, 
chiral symmetry breaking was introduced in the Quark Model. 
This paper continues the program of implementing chiral symmetry
in the Quark Model, showing that the Quark Model also complies 
with some of the most famous PCAC theorems. In particular I address 
the Gell-Mann Oakes and Renner relation
\cite{Gell-Mann}, 
the Goldberger and Treiman Relation
\cite{Goldberger}, 
the Adler self consistency zero
\cite{Adler}
and the Weinberg theorem
\cite{Weinberg}.

\par
I do not aim to derive here new theorems for 
pion physics. 
Since the pioneering work of Yukawa, pion properties 
have already been understood with the techniques of 
Current Algebra, of the Sigma Model, of the Nambu and 
Jona Lasinio model, and of Chiral Lagrangians.
The goal of this paper is to achieve the same perfect
understanding of chiral symmetry breaking in the quark 
model. 
This understanding is not trivial in the Quark Model 
because the pion is an extended \cite{pion size} and 
composite meson, composed of a quark-antiquark pair.
Recently Bjorken asked 
''{\em how are the many disparate methods of describing hadrons 
which are now in use related to each other and to the first 
principles of QCD?}'' 
Here the missing link between the Quark Model and the low energy 
unique field theory of pions is investigated. 
This work clarifies what classes of 
diagrams are necessary to recover the pion theorems in the 
quark model, and explicitly shows the role of the axial Ward identity 
in the Quark Model. This is potentially useful to the
numerous hadronic processes that the Quark Model addresses.

\par
Moreover it is important to stress that the Quark Model 
provides an explicit prescription to address virtual 
pions with off mass shell momenta. The Quark Model is 
suited to describe the virtual exchange of a meson with 
momentum equal to the sum of the quark and antiquark momenta,
and different from the momentum of the mass shell.
The relevant experimental processes that I study
here are the neutron decay and pi-pi scattering.
In neutron decay a virtual pion is produced by the nucleon. 
Moreover the experiments use at least one virtual 
pion in pi-pi scattering because two beams of pions have not yet been 
scattered in the laboratory.
One should acknowledge that it is possible to extract mass shell pi-pi 
scattering parameters from pion-nucleon scattering and form kaon to pion-pion
decay \cite{Colangelo}, 
and that an improvement on data is expected in the new DIRAC 
\cite{Yazov}
experiment at CERN which will soon be able to measure directly pi-pi 
scattering both on the mass shell and at the threshold.
Nevertheless there is also interesting data for pi-pi scattering 
off the mass shell.
For instance the $\pi-\pi$ phase shifts are experimentally 
estimated with the help of $\pi \ N \rightarrow \pi \ \pi \ N$ scattering
\cite{Martin}. In a possible contribution to $\pi \ N \rightarrow \pi \ \pi \ N$
at threshold, the nucleon provides a virtual pion 
$\pi^*$ with offshellness $P^2-M_\pi^2=-3.32 M_\pi^2$, 
see Fig. \ref{offshell} (a). 
Another experiment is $K^+ \rightarrow \pi^+ \ \pi^+ \ \pi^-$ where the
kaon provides a virtual pion with offshellness $P^2-M_\pi^2=+ 10.75 M_\pi^2$, 
see Fig. \ref{offshell} (b).

\par 
The Quark Model is usually understood with simple quantum mechanics. 
Baryons are bound states of three quark, mesons are quark-antiquark 
bound states, and both are studied with the Schr\"odinger equation. 
The hadronic reactions are also studied with coupled channels equations, 
and the couplings are computed with the Resonating Group Method. 
In this paper I choose to display the equations with the compact notation of Feynman 
diagrams, following the simplifying principles of Llewellyn-Smith in his proof of
the Bethe-Salpeter normalization condition 
\cite{Llewellyn-Smith}. 
This decreases the number of terms involved in the equations because the Feynman 
propagator includes both the quark and the antiquark poles,
\begin{equation}
{i \over \not k -m +i \epsilon}
= {i \, \sum_su_su^{\dagger}_s \, \beta \over k_0 -E +i \epsilon} \
 - {i \, \sum_sv_sv^{\dagger}_s \, \beta \over -k_0 -E +i \epsilon} 
\label{quark pole}
\end{equation}
where $u_s({\bf k})$ and $v_s({\bf k})$ are the quark and the antiquark Dirac spinors.
The translation from the covariant Feynman notation to the non-relativistic 
notation {\em is direct and exact}, and is based on eq. (\ref{quark pole}). 
Incidently the formalism of Feynman diagrams
applies straightforwardly to relativistic models like the Nambu and Jona-Lasinio model 
\cite{Nambu,Veronique} 
and other models with Euclidean space integrations
\cite{pi-pi,Liu,Roberts2,Smekal}, 
and also to covariant models in Minkowsky space
\cite{Sauli}. 

\par
The essential simplicity of the Quark Model resides in using {\em only two-body
and finite} quark-antiquark interactions. This is equivalent to use only planar
interactions in the possible series of Feynman diagrams, which are also
obtained in the large $N_c$ (number of colors) limit of QCD
\cite{Veronique}. 
In particular
the intermediate meson exchange is described by the ladder series,
\begin{equation}
\begin{picture}(222,30)(0,0) 
\put(0,10){                         
\begin{picture}(40,20)(0,7)
\put(0,15){\line(1,0){5}}
\put(0,5){\vector(1,0){10}}
\put(15,15){\vector(-1,0){10}}
\put(15,5){\line(-1,0){5}}
\put(15,0){\framebox(10,20){}}
\put(25,15){\line(1,0){5}}
\put(25,5){\vector(1,0){10}}
\put(40,15){\vector(-1,0){10}}
\put(40,5){\line(-1,0){5}}
\end{picture}
$ \ = \ $
\begin{picture}(15,20)(0,7)
\put(0,15){\line(1,0){5}}
\put(0,5){\vector(1,0){10}}
\put(15,15){\vector(-1,0){10}}
\put(15,5){\line(-1,0){5}}
\end{picture}
$ \ + $
\begin{picture}(30,20)(0,7)
\put(0,0){
\begin{picture}(15,20)(0,0)
\put(0,15){\line(1,0){5}}
\put(0,5){\vector(1,0){10}}
\put(15,15){\vector(-1,0){10}}
\put(15,5){\line(-1,0){5}}
\end{picture}}
\put(15,0){
\begin{picture}(15,20)(0,0)
\multiput(0,3)(0,2){6}{$\cdot$}
\put(0,15){\line(1,0){5}}
\put(0,5){\vector(1,0){10}}
\put(15,15){\vector(-1,0){10}}
\put(15,5){\line(-1,0){5}}
\end{picture}}
\end{picture}
$ \ + $
\begin{picture}(45,20)(0,7)
\put(0,0){
\begin{picture}(15,20)(0,0)
\put(0,15){\line(1,0){5}}
\put(0,5){\vector(1,0){10}}
\put(15,15){\vector(-1,0){10}}
\put(15,5){\line(-1,0){5}}
\end{picture}}
\put(15,0){
\begin{picture}(15,20)(0,0)
\multiput(0,3)(0,2){6}{$\cdot$}
\put(0,15){\line(1,0){5}}
\put(0,5){\vector(1,0){10}}
\put(15,15){\vector(-1,0){10}}
\put(15,5){\line(-1,0){5}}
\end{picture}}
\put(30,0){
\begin{picture}(15,20)(0,0)
\multiput(0,3)(0,2){6}{$\cdot$}
\put(0,15){\line(1,0){5}}
\put(0,5){\vector(1,0){10}}
\put(15,15){\vector(-1,0){10}}
\put(15,5){\line(-1,0){5}}
\end{picture}}
\end{picture}
$ \ + \dots  $
}
\end{picture}
\end{equation}
where the dotted line corresponds to the chiral
invariant quark-quark interaction of vertex $V$
and of local kernel $\cal K$. As usual in the Quark Model 
the vertex $V$ is color dependent
and includes a Gell-Mann matrix $\lambda^a / 2$. 
The arrowed line corresponds
to the Feynman quark propagator. In this paper
the direct coupling of three or four mesons is studied. 
I use the technique of dressing the corresponding
Feynman loop with all possible planar insertions  of 
the quark-antiquark interaction, and
to re-sum the obtained series in terms of the 
quark-antiquark ladder. Again, the
ladder is well defined for any total momentum,
and this includes off mass shell momenta.

\par
I also assume that chiral symmetry is spontaneously 
broken in the Quark Model. This is the only assumption in this
paper which goes beyond the minimal Quark Model. 
However the phenomenological success of PCAC shows that it is 
crucial to include chiral symmetry in the Quark Model. 
Therefore the vertex $V$ is assumed to
anticommute with $\gamma_5$. Frequently a vector vertex inspired
in the gluon coupling is used for $V$, but other Dirac structures
for the vertex $V$ can also be also used 
\cite{vector axial}.
Moreover the bare quark propagator
\begin{equation}
S_0(k)= {i  \over \not k + m + i \epsilon }
\label{bare quark}
\end{equation}
must be replaced, in the computed Feynman loops,
by the dressed quark propagator
\begin{equation}
S(k)= {i  \over A(k^2)\not k + B(k^2) + i \epsilon }
\label{dressed quark}
\end{equation}
where the functions $A$ and $B$ are non-trivial solutions
of the mass gap equation and include the scale of the
interaction which is comparable to $\Lambda_{QCD}$. The
current quark mass $m$ is much smaller than the scale 
$\Lambda_{QCD}$, and therefore it only affects pertubatively
$A$ and $B$. In what concerns bound states, the degeneracy 
of chiral partners is broken, in particular
the $\pi$ is a Goldstone boson in the chiral limit. 
These basic properties of the Quark Model with chiral
symmetry have been understood long ago with covariant
\cite{Pagels,Scadron}
quark models with the Schwinger-Dyson equation, 
and some time ago in equal time quark models,
\cite{Pene,Adler,Bicudo,scalar}
with the mass gap equation,
and therefore they are used as a starting point in this paper.
\par
With the concern of deriving a general proof that the Quark Model complies
with the PCAC relations, I follow in this paper the logical path of using the 
simplest PCAC relations as the necessary intermediate steps to arrive at the
rather technical proof of the Weinberg Theorem for $\pi-\pi$ scattering.
The Sections II and III define the formalism of this paper.
This formalism is standard, nevertheless it is convenient to
define it clearly.
Section II reviews mesons as quark-antiquark bound states in 
the Ladder framework. 
Section III reviews the Axial Ward Identity which is crucial 
for the low energy pion theorems. 
Sections IV and V apply the techniques defined in
Sections II and III to standard PCAC relations,
which have been extensively studied in the literature. 
This checks the methods used in this paper.
Section IV recovers the Gell-Mann, Oakes and Renner 
relation. Section V recovers the Goldberger and Treiman relation. 
Once the formalism is defined and checked, Sections VI and VII
explicitly study the more technical PCAC relations.
Section VI proves that the Quark Models possess 
the Adler Self-consistency Zeroes. 
Section VII proves that the Quark Models 
comply with the Weinberg Theorem. 
The conclusion is presented in Section VIII.

\section{Quarks, mesons and the Ladder} 

\par
The ladder series is a geometrical series which
includes bound states. A meson is a quark-antiquark 
bound state which corresponds to a pole in the 
series. Outside the pole the ladder does not describe
asymptotic states, nevertheless ladder exchange
appears as a sub diagram contributing to the interaction
of asymptotic states. Then the ladder includes both
the off mass shell exchange of mesons, and
the contact interaction term. 

\par
I follow the usual convention of factorizing the
pole and the Bethe-Salpeter vertices. In the
close neighborhood of a bound state 
$b$, a pole $M^2_b$ occurs in the external momentum $P^2$,
and the ladder obeys the spectral decomposition,
\FL
\begin{equation}
\begin{picture}(47,20)(0,0)
\put(0,-8){
%
\put(1,5){
\begin{picture}(40,20)(0,0)
\put(0,15){\line(1,0){5}}
\put(0,5){\vector(1,0){10}}
\put(15,15){\vector(-1,0){10}}
\put(15,5){\line(-1,0){5}}
\put(15,0){\framebox(10,20){}}
\put(25,15){\line(1,0){5}}
\put(25,5){\vector(1,0){10}}
\put(40,15){\vector(-1,0){10}}
\put(40,5){\line(-1,0){5}}
\put(17,-7){\vector(-1,0){10}}
\put(20,-10){$P$}
\end{picture}}
}
\end{picture}
=
\begin{picture}(42,15)(0,0)
\put(-5,-10){
%
\put(-20,5){
\begin{picture}(40,20)(0,0)
\put(25,15){\line(1,0){5}}
\put(25,5){\vector(1,0){10}}
\put(40,15){\vector(-1,0){10}}
\put(40,5){\line(-1,0){5}}
\end{picture}}
\put(-20,5){
\begin{picture}(40,20)(0,0)
\put(40,10){\oval(10,10)[r]}
\put(43,8){$\bullet$}
\put(48,10){${\chi_b}_{_{P}}$}
\end{picture}}
}
\end{picture}
\
{ i \over P^2-M_b^2 +i \epsilon} 
\
\begin{picture}(51,15)(0,0)
\put(-4,-10){
%
\put(32,5){
\begin{picture}(40,20)(0,0)
\put(0,15){\line(1,0){5}}
\put(0,5){\vector(1,0){10}}
\put(15,15){\vector(-1,0){10}}
\put(15,5){\line(-1,0){5}}
\end{picture}}
\put(-13,5){
\begin{picture}(40,20)(0,0)
\put(45,10){\oval(10,10)[l]}
\put(37,8){$\bullet$}
\put(15,10){${\chi_b}_{_{-P}} $}
\end{picture}}
}
\end{picture}
\label{ladder pole}
\end{equation}
where
\begin{picture}(25,10)(0,0)
\put(3,2.5){$\bullet$}
\put(8,5){${\chi_b}_P$}
\put(0,5){\oval(10,10)[r]}
\end{picture}
is  the Bethe-Salpeter vertex, or truncated
amplitude, of a meson, and the arrowed 
\begin{picture}(15,2)(0,0)
\put(15,1){\vector(-1,0){10}}
\put(5,1){\line(-1,0){5}}
\end{picture}
line represents
a dressed quark propagator $\cal S$. The non amputated amplitude is
simply obtained with the product  ${\cal S} {\chi_b}_P {\cal S}$.
Eq. (\ref{ladder pole}) and the rest of
the paper follows the convention where the momentum $P^\mu$ of the
vertex flows inside the quark loop, summing to the outgoing quark line. 
${\chi_b}_{_{-P}} $ has the opposite total momentum, in
particular $-P^0$ is negative.  ${\chi_b}_{_{-P}} $ 
can also  be obtained from 
${\chi_b}_P$ with the charge conjugation transformation. 
${\chi_b}_P$ is a function of the relative momentum
$k$ of the bound pair of a quark and an antiquark. 

\par
In what concerns
the total four-momentum $P^\mu$, the bound state vertex is straightforwardly 
defined in the mass shell, which corresponds to the exact momentum of the
pole $P^2=M_b^2$. Nevertheless with eq.(\ref{ladder pole}) it is possible 
to extend the definition of  ${\chi_b}_P$ to a small neighborhood of
the pole, up to first order in the off mass shell quantity
$P^2-M_b^2$. 
In a diagrammatic language, {\em the ladder includes 
both the exchange of mesons and the contact interaction}. 
The off mass shell vertex also includes the principal part
of the ladder series. The principal part contains the contact interaction,
and also contains the infinite tower of excited states which are
a solution of the bound state equation when the potential is confining.
Extending the Bethe Salpeter vertex off the mass shell constitutes an
economical method to include all these effects. The off mass shell
vertex is well defined at least in a small neighborhood of the pole.

\par
To compute the Bethe-Salpeter vertex, it is convenient to
rewrite the ladder in a self-consistent equation,
%
%
\begin{equation}
%
\begin{picture}(50,20)(0,0) 
\put(0,-5){
%
\put(0,5){                         
\begin{picture}(25,20)(0,0)
\put(0,15){\line(1,0){5}}
\put(0,5){\vector(1,0){10}}
\put(15,15){\vector(-1,0){10}}
\put(15,5){\line(-1,0){5}}
\put(15,0){\framebox(10,20){}}
\end{picture}}
\put(25,5){                         
\begin{picture}(15,20)(0,0)
\put(0,15){\line(1,0){5}}
\put(0,5){\vector(1,0){10}}
\put(15,15){\vector(-1,0){10}}
\put(15,5){\line(-1,0){5}}
\end{picture}}
}
\end{picture}
=
\begin{picture}(25,15)(0,0) 
\put(0,-5){
%
\put(0,5){                         
\begin{picture}(15,20)(0,0)
\put(0,15){\line(1,0){5}}
\put(0,5){\vector(1,0){10}}
\put(15,15){\vector(-1,0){10}}
\put(15,5){\line(-1,0){5}}
\end{picture}}
}
\end{picture}
+
\begin{picture}(65,20)(0,0) 
\put(0,-5){
\put(0,5){                         
\begin{picture}(15,20)(0,0)
\multiput(15,3)(0,2){6}{$\cdot$}
\put(0,15){\line(1,0){5}}
\put(0,5){\vector(1,0){10}}
\put(15,15){\vector(-1,0){10}}
\put(15,5){\line(-1,0){5}}
\end{picture}}
\put(15,5){                         
\begin{picture}(25,20)(0,0)
\put(0,15){\line(1,0){5}}
\put(0,5){\vector(1,0){10}}
\put(15,15){\vector(-1,0){10}}
\put(15,5){\line(-1,0){5}}
\put(15,0){\framebox(10,20){}}
\end{picture}}
\put(40,5){
\begin{picture}(15,20)(0,0)
\put(0,15){\line(1,0){5}}
\put(0,5){\vector(1,0){10}}
\put(15,15){\vector(-1,0){10}}
\put(15,5){\line(-1,0){5}}
\end{picture}}
}
\end{picture}
%
\ .
\label{ladder}
\end{equation}
Replacing eq. (\ref{ladder pole}) in eq. (\ref{ladder}), 
and folding it from the left with $\chi_{_P}$,
the off mass shell Bethe Salpeter 
equation is obtained,
\begin{eqnarray}
\label{off mass shell}
\begin{picture}(20,20)(0,0) 
\put(0,0){
%
\put(0,-5){
\begin{picture}(10,20)(0,0)
\put(3,7.5){$\bullet$}
\put(8,10){$\chi_{_P}$}
\put(0,10){\oval(10,10)[r]}
\end{picture}}
}
\end{picture}
&=&
\begin{picture}(40,15)(0,0) 
\put(0,-10){
%
\put(0,5){                         
\begin{picture}(15,20)(0,0)
\multiput(0,3)(0,2){6}{$\cdot$}
\put(0,15){\line(1,0){5}}
\put(0,5){\vector(1,0){10}}
\put(15,15){\vector(-1,0){10}}
\put(15,5){\line(-1,0){5}}
\end{picture}}
\put(15,5){
\begin{picture}(10,20)(0,0)
\put(0,10){\oval(10,10)[r]}
\put(3,7.5){$\bullet$}
\put(8,10){$\chi_{_P}$}
\end{picture}}
}
\end{picture}
\left( 1 - { P^2-M^2 \over i \, {\cal I} } \right)^{-1}
\ ,
\\
{\cal I} &=&
\begin{picture}(65,20)(0,0) 
\put(0,-10){
%
\put(0,5){
\begin{picture}(15,20)(0,0)
\put(25,10){\oval(10,10)[l]}
\put(17,7.5){$\bullet$}
\put(0,10){$\chi_{_{-P}}$}
\end{picture}}
\put(25,5){                         
\begin{picture}(15,20)(0,0)
\put(0,15){\line(1,0){5}}
\put(0,5){\vector(1,0){10}}
\put(15,15){\vector(-1,0){10}}
\put(15,5){\line(-1,0){5}}
\end{picture}}
\put(40,5){
\begin{picture}(10,20)(0,0)
\put(0,10){\oval(10,10)[r]}
\put(3,7.5){$\bullet$}
\put(8,10){$\chi_{_P}$}
\end{picture}}
}
\end{picture}
\nonumber \\
&=&\int tr \bigl\{ \chi_{_P}(k){\cal S}(k+{P/2})\chi_{_{-P}}(k){\cal S}(k-{P/2})\bigr\} 
\ ,
\nonumber
\label{off mass shell}
\end{eqnarray}
where ${\cal I}$ is both displayed as a Feynman loop and as an integral. 
For compactness, the convention of representing integrals of propagators and 
vertices with Feynman Diagrams will mainly be used in the rest of the paper.
The loop ${\cal I}$ is finite and proportional 
to the square of the scale of the interaction. 
Nevertheless ${\cal I}$ will factorize from the results 
of this paper, which are model independent. 
At the mass shell momentum ,
eq. (\ref{off mass shell}) simplifies to the standard
Bethe Salpeter equation,
\begin{equation}
\begin{picture}(20,20)(0,0) 
\put(0,0){
%
\put(0,-5){
\begin{picture}(10,20)(0,0)
\put(3,7.5){$\bullet$}
\put(8,10){$\chi_P$}
\put(0,10){\oval(10,10)[r]}
\end{picture}}
}
\end{picture}
=
\begin{picture}(40,15)(0,0) 
\put(0,-10){
%
\put(0,5){                         
\begin{picture}(15,20)(0,0)
\multiput(0,3)(0,2){6}{$\cdot$}
\put(0,15){\line(1,0){5}}
\put(0,5){\vector(1,0){10}}
\put(15,15){\vector(-1,0){10}}
\put(15,5){\line(-1,0){5}}
\end{picture}}
\put(15,5){
\begin{picture}(10,20)(0,0)
\put(0,10){\oval(10,10)[r]}
\put(3,7.5){$\bullet$}
\put(8,10){$\chi_P$}
\end{picture}}
} 
\end{picture}
\ .
\label{Bethe Salpeter}
\end{equation}

\par
To check that the off mass shell eq. (\ref{off mass shell}) Bethe Salpeter 
equation is correct, I derive from it the
normalization condition
\cite{Llewellyn-Smith} for the Bethe-Salpeter vertices. 
Folding from the right with 
$\chi_{_{-P}}$, eq. (\ref{off mass shell}) becomes,  
\FL
\begin{equation}
%
%
\begin{picture}(75,20)(0,0) 
\put(0,0){
%
\put(-2,0){
\begin{picture}(15,20)(0,0)
\put(25,10){\oval(10,10)[l]}
\put(17,7.5){$\bullet$}
\put(0,10){$\chi_{_{-P}}$}
\end{picture}}
\put(13,0){                         
\begin{picture}(15,20)(0,0)
\put(10,15){\line(1,0){5}}
\put(10,5){\vector(1,0){10}}
\put(25,15){\vector(-1,0){10}}
\put(25,5){\line(-1,0){5}}
\multiput(23,3)(0,2){6}{$\cdot$}
\end{picture}}
\put(28,0){                         
\begin{picture}(15,20)(0,0)
\put(10,15){\line(1,0){5}}
\put(10,5){\vector(1,0){10}}
\put(25,15){\vector(-1,0){10}}
\put(25,5){\line(-1,0){5}}
\end{picture}}
\put(53,0){
\begin{picture}(10,20)(0,0)
\put(3,7.5){$\bullet$}
\put(8,10){$\chi_{_P}$}
\put(0,10){\oval(10,10)[r]}
\end{picture}}
}
\end{picture}
-
\begin{picture}(60,20)(0,0) 
\put(0,0){
%
\put(-2,0){
\begin{picture}(15,20)(0,0)
\put(25,10){\oval(10,10)[l]}
\put(17,7.5){$\bullet$}
\put(0,10){$\chi_{_{-P}}$}
\end{picture}}
\put(13,0){                         
\begin{picture}(15,20)(0,0)
\put(10,15){\line(1,0){5}}
\put(10,5){\vector(1,0){10}}
\put(25,15){\vector(-1,0){10}}
\put(25,5){\line(-1,0){5}}
\end{picture}}
\put(38,0){
\begin{picture}(10,20)(0,0)
\put(3,7.5){$\bullet$}
\put(8,10){$\chi_{_P}$}
\put(0,10){\oval(10,10)[r]}
\end{picture}}
}
\end{picture}
=i\left(P^2-M^2 \right) 
\label{intermediate norm}
\end{equation}
and the correct normalizing condition 
is obtained when
eq. (\ref{intermediate norm}) is derived by $\partial / \partial P^\mu$. The
derivative of the left hand side is,
\FL
\begin{eqnarray}
&&
%
%
%
\begin{picture}(160,20)(0,0) 
\put(0,0){
%
\put(-2,0){                         
\begin{picture}(25,20)(0,0)
\put(5,5 ){${\partial \over \partial_{P^\mu}}$}
\put(25,7){$\Bigl($}
\put(70,7){$\Bigr)$}
\end{picture}}
\put(28,0){
\begin{picture}(15,20)(0,0)
\put(25,10){\oval(10,10)[l]}
\put(17,7.5){$\bullet$}
\put(0,10){$\chi_{_{-P}}$}
\end{picture}}
\put(43,0){                         
\begin{picture}(15,20)(0,0)
\put(10,15){\line(1,0){5}}
\put(10,5){\vector(1,0){10}}
\put(25,15){\vector(-1,0){10}}
\put(25,5){\line(-1,0){5}}
\end{picture}}
\put(78,0){                         
\begin{picture}(25,20)(0,0)
\put(0,7){$\Bigl($}
\put(70,7){$\Bigr)$}
\end{picture}}
\put(75,0){                         
\begin{picture}(15,20)(0,0)
\multiput(8,3)(0,2){6}{$\cdot$}
\put(10,15){\line(1,0){5}}
\put(10,5){\vector(1,0){10}}
\put(25,15){\vector(-1,0){10}}
\put(25,5){\line(-1,0){5}}
\end{picture}}
\put(100,0){
\begin{picture}(10,20)(0,0)
\put(3,7.5){$\bullet$}
\put(8,10){$\chi_{_P}$}
\put(0,10){\oval(10,10)[r]}
\end{picture}}
\put(130,0){
\begin{picture}(10,20)(0,0)
\put(-10,7.5){$-$}
\put(3,7.5){$\bullet$}
\put(8,10){$\chi_{_P}$}
\put(0,10){\oval(10,10)[r]}
\end{picture}}
}
\end{picture}
%
+
\\
\nonumber 
&&
%
%
\begin{picture}(185,20)(0,0) 
\put(0,0){
%
\put(3,0){                         
\begin{picture}(25,20)(0,0)
\put(0,7){$\Bigl($}
\put(90,7){$\Bigr)$}
\end{picture}}
\put(8,0){
\begin{picture}(15,20)(0,0)
\put(25,10){\oval(10,10)[l]}
\put(17,7.5){$\bullet$}
\put(0,10){$\chi_{_{-P}}$}
\end{picture}}
\put(23,0){                         
\begin{picture}(15,20)(0,0)
\put(10,15){\line(1,0){5}}
\put(10,5){\vector(1,0){10}}
\put(25,15){\vector(-1,0){10}}
\put(25,5){\line(-1,0){5}}
\multiput(23,3)(0,2){6}{$\cdot$}
\end{picture}}
\put(65,0){
\begin{picture}(10,20)(0,0)
\put(-10,7.5){$-$}
\put(25,10){\oval(10,10)[l]}
\put(17,7.5){$\bullet$}
\put(0,10){$\chi_{_{-P}}$}
\end{picture}}
\put(100,0){                         
\begin{picture}(25,20)(0,0)
\put(5,5 ){${\partial \over \partial_{P^\mu}}$}
\put(25,7){$\Bigl($}
\put(70,7){$\Bigr)$}
\end{picture}}
\put(125,0){                         
\begin{picture}(15,20)(0,0)
\put(10,15){\line(1,0){5}}
\put(10,5){\vector(1,0){10}}
\put(25,15){\vector(-1,0){10}}
\put(25,5){\line(-1,0){5}}
\end{picture}}
\put(150,0){
\begin{picture}(10,20)(0,0)
\put(3,7.5){$\bullet$}
\put(8,10){$\chi_{_P}$}
\put(0,10){\oval(10,10)[r]}
\end{picture}}
}
\end{picture}
+
\\
\nonumber 
&&
%
%
%
%
\begin{picture}(122,20)(0,0) 
\put(-2,0){
\put(0,0){
\begin{picture}(10,20)(0,0)
\put(25,10){\oval(10,10)[l]}
\put(17,7.5){$\bullet$}
\put(0,10){$\chi_{_{-P}}$}
\end{picture}}
\put(15,0){                         
\begin{picture}(10,20)(0,0)
\put(10,15){\line(1,0){5}}
\put(10,5){\vector(1,0){10}}
\put(25,15){\vector(-1,0){10}}
\put(25,5){\line(-1,0){5}}
\end{picture}}
\put(40,0){                         
\begin{picture}(25,20)(0,0)
\put(5,5 ){${\partial \over \partial_{P^\mu}}$}
\put(22,7){$\Bigl($}
\put(37,7){$\Bigr)$}
\end{picture}}
\put(60,0){                         
\begin{picture}(15,20)(0,0)
\put(10,15){\line(1,0){5}}
\put(10,5){\line(1,0){5}}
\multiput(12,3)(0,2){6}{$\cdot$}
\end{picture}}
\put(75,0){                         
\begin{picture}(15,20)(0,0)
\put(10,15){\line(1,0){5}}
\put(10,5){\vector(1,0){10}}
\put(25,15){\vector(-1,0){10}}
\put(25,5){\line(-1,0){5}}
\end{picture}}
\put(100,0){
\begin{picture}(10,20)(0,0)
\put(3,7.5){$\bullet$}
\put(8,10){$\chi_{_P}$}
\put(0,10){\oval(10,10)[r]}
\end{picture}}
}
\end{picture}
+
%
%
\begin{picture}(105,20)(0,0) 
\put(-2,0){
\put(0,0){
\begin{picture}(15,20)(0,0)
\put(20,10){\oval(10,10)[l]}
\put(12,7.5){$\bullet$}
\put(0,10){$\chi_P$}
\end{picture}}
\put(20,0){                         
\begin{picture}(25,20)(0,0)
\put(5,5 ){${\partial \over \partial_{P^\mu}}$}
\put(22,7){$\Bigl($}
\put(49,7){$\Bigr)$}
\end{picture}}
\put(40,0){                         
\begin{picture}(15,20)(0,0)
\put(10,15){\line(1,0){5}}
\put(10,5){\vector(1,0){10}}
\put(25,15){\vector(-1,0){10}}
\put(25,5){\line(-1,0){5}}
\end{picture}}
\put(80,0){
\begin{picture}(10,20)(0,0)
\put(3,7.5){$\bullet$}
\put(8,10){$\chi_P$}
\put(0,10){\oval(10,10)[r]}
\end{picture}}
}
\end{picture}
%
%
\end{eqnarray}
and this provides a general normalizing condition for the vertex.
Frequently mass shell vertices and local kernels are used.
Then the Bethe Salpeter (\ref{Bethe Salpeter})
equation can be used to precisely cancel the terms with the derivative of the vertices 
$\chi_P$ and $\chi_{-P}$, and the derivative of the kernel also vanishes 
because the kernel (the quark-quark interaction) is local. With these cancellations,
the derivative of eq. (\ref{intermediate norm}) is simply, 
\begin{equation}
\begin{picture}(110,20)(0,0) 
\put(0,0){
%
\put(0,0){
\begin{picture}(15,20)(0,0)
\put(20,10){\oval(10,10)[l]}
\put(12,7.5){$\bullet$}
\put(0,10){$\chi_P$}
\end{picture}}
\put(20,0){                         
\begin{picture}(25,20)(0,0)
\put(5,5 ){${\partial \over \partial_{P^\mu}}$}
\put(25,7){$\Bigl($}
\put(55,7){$\Bigr)$}
\end{picture}}
\put(45,0){                         
\begin{picture}(15,20)(0,0)
\put(10,15){\line(1,0){5}}
\put(10,5){\vector(1,0){10}}
\put(25,15){\vector(-1,0){10}}
\put(25,5){\line(-1,0){5}}
\end{picture}}
\put(85,0){
\begin{picture}(10,20)(0,0)
\put(3,7.5){$\bullet$}
\put(8,10){$\chi_P$}
\put(0,10){\oval(10,10)[r]}
\end{picture}}
}
\end{picture}
= 2 \, i \, P_\mu
\ .
\label{normalizing}
\end{equation}
This is the standard normalizing condition for mass shell Bethe Salpeter vertices 
with a local kernel.

\par
The off mass shell eq. (\ref{off mass shell}) is particularly
simple in the case where the bound state is a low energy pion, 
In this case the expansion in the external $P^\mu$ and in $M_\pi$
can be used, because $M_\pi$ and $P^\mu$ are much smaller that the 
characteristic scale of meson physics, say $2\Lambda _{QCD}$.
The off mass shell correction only starts contributing to the Bethe
Salpeter equation (\ref{off mass shell}) at the second order of $P^2$ 
and $M^2$. 
Therefore, up to the first order in $P^\mu$ and in $M_\pi$ , 
the vertex $\chi_{_P}$ is {\em formally the same function } of 
$P^\mu$, both for mass shell and for off mass shell pions. 
For instance the momentum
expansion up to first order in $P^\mu$ of the pion Bethe Salpeter vertex,
\begin{eqnarray}
\chi_P(k)&&=\chi^0(k)+P^\mu \,\chi^1_\mu(k) + o(P^\mu\,P^\nu) \ ,
\label{vertex and momentum}
\\
&&\chi^1_\mu(k)=\Bigl( F(k) \gamma_\mu 
+ G(k) k_\mu \not k + H(k) [ \gamma_\mu , \not k ] \Bigr) \gamma_5 
\nonumber 
\end{eqnarray}  
is also correct, and formally the same, for any small off the mass shell
momentum $P^\mu$. In particular the expansion in $P^\mu$
of the Bethe Salpeter equation (\ref{Bethe Salpeter}) yields for
$\chi^0(k)$ and for the components of $\chi^1_\mu(k)$ four equations totally 
independent of $P^\mu$. $\chi^0(k)$ will be exactly derived in eq. (\ref{chi = ga}).
Importantly, the four components of $\chi_\pi$ will only contribute to 
the PCAC theorems of this paper
through the pion decay constant $f_\pi$, which is defined by the trace,
\begin{equation}
tr\{ (S \chi S)_{_{P}}
\gamma^\mu  \gamma^5 \} = \sqrt{2} f_\pi P^\mu \ ,
\label{decay}
\end{equation}
where $ \sqrt{2} $ is a flavor factor.
In eq. (\ref{decay}) and in the rest of the
paper the traces are assumed to include the momentum integral and the sum
in Dirac and color indices.
The important result of this low energy pion discussion is that
(\ref{decay}) is also correct outside the mass shell for $P^2 \neq M^2$, when
a virtual intermediate pion is used, providing the momentum $P^\mu$ is small.
 
\par
I now derive a second important relation, which states how to include 
(or remove) a ladder in the vertex $\chi_{_P}$. 
Folding eq. (\ref{ladder pole}) from the right with the vertex $\chi_{_P}$, 
and dividing by the $\cal I$ loop,
%
%
\begin{equation}
\begin{picture}(35,15)(0,0) 
\put(0,-10){
%
\put(0,5){                         
\begin{picture}(15,20)(0,0)
\put(0,15){\line(1,0){5}}
\put(0,5){\vector(1,0){10}}
\put(15,15){\vector(-1,0){10}}
\put(15,5){\line(-1,0){5}}
\end{picture}}
\put(15,5){
\begin{picture}(10,20)(0,0)
\put(0,10){\oval(10,10)[r]}
\put(3,7.5){$\bullet$}
\put(8,10){$\chi$}
\end{picture}}
}
\end{picture}
=
\begin{picture}(60,20)(0,0) 
\put(0,-10){
%
\put(0,5){                         
\begin{picture}(25,20)(0,0)
\put(0,15){\line(1,0){5}}
\put(0,5){\vector(1,0){10}}
\put(15,15){\vector(-1,0){10}}
\put(15,5){\line(-1,0){5}}
\put(15,0){\framebox(10,20){}}
\end{picture}}
\put(25,5){                         
\begin{picture}(15,20)(0,0)
\put(0,15){\line(1,0){5}}
\put(0,5){\vector(1,0){10}}
\put(15,15){\vector(-1,0){10}}
\put(15,5){\line(-1,0){5}}
\end{picture}}
\put(40,5){
\begin{picture}(10,20)(0,0)
\put(0,10){\oval(10,10)[r]}
\put(3,7.5){$\bullet$}
\put(8,10){$\chi$}
\end{picture}}
}
\end{picture}
\ { P^2-M^2 \over i \, {\cal I} } 
\ .
\label{pole pion}
\end{equation}

\section{Using the Axial Ward Identity} 

\par
When chiral symmetry breaking occurs, the mass
gap equation has a non-trivial solution.
The Schwinger-Dyson equation for the full propagator is,
\begin{equation}
S(k)^{-1}=S_0^{-1}(k)+i\int {d^4 q \over (2 \pi)^4}
{\cal K}(q) V S(k+q) V  \ . 
\label{SD}
\label{mass gap}
\end{equation}
This equation (\ref{SD}) is also known as the mass gap
equation because the initially almost massless
gap between the quark and antiquark dispersion relations
is increased when the constituent mass $M=\sqrt{B^2 / A^2}$
is generated. Because the vertex $V$ includes the Gell-Mann
matrices, the tadpole does not contribute to eq. (\ref{SD}).    
Multiplying eq. (\ref{SD}) right or left with $\gamma_5$
and summing leads to,  
\begin{eqnarray}
& S(k_1)^{-1}\gamma_5 + \gamma_5 S(k_2)^{-1} 
= S_0(k_1)^{-1}\gamma_5 + \gamma_5 S_0(k_2)^{-1} &
\nonumber \\  
& -i \int {\cal K}(q) V ( S(k_1+q)\gamma_5 + \gamma_5 S(k_2+q) ) V \ , & 
\end{eqnarray}
which is the Bethe Salpeter equation for the vertex,
\FL
\begin{eqnarray}
\Gamma_{\hspace{-.08cm}A}(k_1,k_2) &=& \hspace{-.08cm} 
\gamma_{\hspace{-.08cm}A}(k_1,k_2)
-i\hspace{-.08cm} \int  {d^4 q \over (2 \pi)^4} {\cal K}(q) V S(k_1+q)
\nonumber \\ && 
\Gamma_{\hspace{-.08cm}A} (k_1+q,k_2+q)
S(k_2+q) V \ ,
\label{BS vertex}
\end{eqnarray}
and this shows 
\cite{Scadron,Adler2,vector axial,Maris,scalar}
that the ladder approximation for the bound state is consistent 
with the quark self energy equation in the rainbow approximation.
Both approximations are equivalent to the planar diagram expansion which
is characteristic of the Quark Model.

\par
In the Bethe Salpeter equation (\ref{BS vertex}), the bare and 
dressed vertices are respectively defined
with the same Axial Ward Identity,
\begin{eqnarray}
\Gamma_{\hspace{-.08cm}A}(k_1,k_2) &=& S^{-1}(k_1) \gamma_5 + \gamma_5 S^{-1}(k_2) \ ,
\nonumber \\
\gamma_{\hspace{-.08cm}A}(k_1,k_2)&=&  S^{-1}_0(k_1) \gamma_5 
+ \gamma_5 S^{-1}_0(k_2) \ .
\label{axialWI}
\end{eqnarray}
At this point it is important to clarify that   
in the chiral limit of $m=0$, the bare vertex $\gamma_{\hspace{-.08cm}A}$
is essentially the momentum contracted with the bare axial vertex
$\gamma^\mu \gamma_5$, and in the limit of vanishing momentum
$P_\mu$, the vertex $\gamma_{\hspace{-.08cm}A}$
is essentially the current quark mass times the bare pseudoscalar vertex
$\gamma_5$. In general $\gamma_{\hspace{-.08cm}A}$ is a combination of the 
axial vertex and of the pseudoscalar vertex. 
In what concerns the dressed vertex $\Gamma_{\hspace{-.08cm}A}$, it 
will be used up to second order in the total momentum, and in general the Dirac 
structure of $\Gamma_{\hspace{-.08cm}A}$ has four components, similar in 
structure to the four components of the pion Bethe-Salpeter
vertex (\ref{vertex and momentum}). Therefore this vertex cannot be reduced,
in the Quark Model framework, neither to a pure pseudoscalar term nor to a pure axial
vector term. Nevertheless in the rest
of this paper, for simplicity and because they are defined with the 
axial Ward identity (\ref{axialWI}), $\gamma_{\hspace{-.08cm}A}$ 
will be called the bare axial vertex and $\Gamma_{\hspace{-.08cm}A}$
will be called the dressed axial vertex, although they possess a more 
general Dirac structure.

\par
The bare axial vertex $\gamma_{\hspace{-.08cm}A}$
is computed from the bare quark propagator
(\ref{bare quark}),
\begin{equation}
{\gamma_{\hspace{-.08cm}A}}_P = 
{ ( \not P  - 2 m )\over i} \gamma_5 \ , \ \
P=k_1-k_2 \ .
\label{bare}
\end{equation} 
$\gamma_{\hspace{-.08cm}A}$ is the particular part of the
Bethe Salpeter equation for the vertex (\ref{BS vertex}), and it
vanishes when the current quark mass $m$ is small (chiral limit) and 
at the same time the total momentum $P^\mu$ of the vertex is small.
On the other hand the dressed vertex $\Gamma_{\hspace{-.08cm}A}$ is 
computed from the dressed quark propagator (\ref{dressed quark}),
\FL
\begin{equation}
\Gamma_{\hspace{-.08cm}A}(k_1,k_2) \hspace{-.08cm} = \hspace{-.08cm} 
{  A(k_1) \hspace{-.1cm} 
\not k_1 \hspace{-.05cm} - \hspace{-.05cm} A(k_2) \hspace{-.1cm} \not k_2 \hspace{-.05cm}
- \hspace{-.05cm} B(k_1) \hspace{-.05cm} - \hspace{-.05cm} B(k_2) 
\over i} \gamma_5 . \hspace{-.3cm}
\end{equation}
$\Gamma_{\hspace{-.08cm}A}$ is finite providing spontaneous chiral symmetry 
breaking occurs in eq. (\ref{SD})
to generate a dynamical mass in the dressed quark propagator. 
For instance when the total momentum $P=k_1-k_2$  of the vertex vanishes,
the vertex is simply identical to $2 i \, B(k) \gamma_5$, where $B(k)$ 
is a finite solution of the mass gap equation. 

\par
For simplicity the flavor is not yet included. Flavor will only
be explicitly included at the end of subsection VII. The isoscalar 
axial Ward identity must include the Axial anomaly, which is crucial to the $U(1)$
problem. Nevertheless the pion is an isovector, and in the coupling of a pion 
I do not need to concern with the Axial anomaly. 

\par
I now derive two powerful relations which involve the axial vertices and the
ladder. After iterating the Bethe Salpeter equation (\ref{BS vertex}) for the
dressed axial vertex $\Gamma_{\hspace{-.08cm}A}$, and including the external
propagators, a first useful relation is derived,
%
%
\begin{equation}
S \, \Gamma_{\hspace{-.08cm}A} \, S
= 
\begin{picture}(70,15)(0,0)
\put(0,0){
\put(5,0){
\begin{picture}(40,20)(0,7)
\put(0,15){\line(1,0){5}}
\put(0,5){\vector(1,0){10}}
\put(15,15){\vector(-1,0){10}}
\put(15,5){\line(-1,0){5}}
\put(15,0){\framebox(10,20){}}
\put(25,15){\line(1,0){5}}
\put(25,5){\vector(1,0){10}}
\put(40,15){\vector(-1,0){10}}
\put(40,5){\line(-1,0){5}}
\put(40,10){\oval(10,10)[r]}
\put(43,8){$\bullet$}
\put(49,10){$\gamma_{\hspace{-.08cm}A}$}
\end{picture}
}
}
\end{picture}
\ .
\label{second}
\end{equation}
To derive some of the PCAC proofs it is crucial to use a
second relation which is an extension of eq. (\ref{axialWI}),
%
%
\begin{eqnarray}
\label{crucial}
\begin{picture}(86,30)(0,0)
\put(0,-8){
%
\put(3,10){\begin{picture}(40,20)(0,0)
\put(0,15){\line(1,0){5}}
\put(0,5){\vector(1,0){10}}
\put(15,15){\vector(-1,0){10}}
\put(15,5){\line(-1,0){5}}
\put(15,0){\framebox(10,20){}}
\put(25,15){\line(1,0){5}}
\put(25,5){\vector(1,0){10}}
\put(40,15){\vector(-1,0){10}}
\put(40,5){\line(-1,0){5}}
\end{picture}}
%
\put(3,10){\begin{picture}(60,20)(0,0)
\put(38,2){$\bullet$}
\put(35,-8){$S^{-1}$}
\put(38,13){$\bullet$}
\put(37,20){$\Gamma_{\hspace{-.08cm}A}$}
\end{picture}}
%
\put(43,10){\begin{picture}(40,20)(0,0)
\put(0,15){\line(1,0){5}}
\put(0,5){\vector(1,0){10}}
\put(15,15){\vector(-1,0){10}}
\put(15,5){\line(-1,0){5}}
\put(15,0){\framebox(10,20){}}
\put(25,15){\line(1,0){5}}
\put(25,5){\vector(1,0){10}}
\put(40,15){\vector(-1,0){10}}
\put(40,5){\line(-1,0){5}}
\end{picture}}
}
\end{picture}
&=&
\begin{picture}(56,30)(0,0)
\put(0,-8){
\put(13,5){\begin{picture}(40,20)(0,0)
\put(0,15){\line(1,0){5}}
\put(0,5){\vector(1,0){10}}
\put(15,15){\vector(-1,0){10}}
\put(15,5){\line(-1,0){5}}
\put(15,0){\framebox(10,20){}}
\put(25,15){\line(1,0){5}}
\put(25,5){\vector(1,0){10}}
\put(40,15){\vector(-1,0){10}}
\put(40,5){\line(-1,0){5}}
\put(-11,15){$\gamma_5 $}
\end{picture}}
}
\end{picture}
+
\begin{picture}(54,30)(0,0)
\put(0,-8){
\put(3,5){\begin{picture}(40,20)(0,0)
\put(0,15){\line(1,0){5}}
\put(0,5){\vector(1,0){10}}
\put(15,15){\vector(-1,0){10}}
\put(15,5){\line(-1,0){5}}
\put(15,0){\framebox(10,20){}}
\put(25,15){\line(1,0){5}}
\put(25,5){\vector(1,0){10}}
\put(40,15){\vector(-1,0){10}}
\put(40,5){\line(-1,0){5}}
\put(40,15){$\gamma_5 $}
\end{picture}}
}
\end{picture}
\end{eqnarray}
and this constitutes a Ward identity for the ladder.
This identity is derived if I expand 
\cite{scalar}
the ladders and 
substitute the vertex in the left hand side. Then all terms with 
an intermediate $\gamma_5$ include the anticommutator $\{\gamma_5,V\}$
and this cancels because {\em the interaction is chiral
invariant} and {\em the kernel is local}. 
Only the right hand side survives.

\section{The Gell-Mann Oakes and Renner relation} 

\par
In the limit of vanishing current quark mass $m$ and 
vertex momentum $P^\mu$, the Bethe-Salpeter (\ref{BS vertex})
equation for the 
axial vertex $\Gamma_A$ becomes homogeneous and is
thus identical to a homogeneous 
Bethe-Salpeter equation (\ref{Bethe Salpeter}) for a
pion vertex ${\chi_\pi}_P(k)$ with vanishing mass. 
In this limit the pion is a massless Goldstone boson, and 
the pion Bethe Salpeter vertex is proportional to the dressed 
axial vertex, and to the dynamical quark mass $B(k)$,
\begin{equation}
\chi_{_0}(k)= { B(k) \over n_\pi }\gamma_5 
= {1 \over 2 \, i \, n_\pi } {\Gamma_{\hspace{-.08cm}A}}_0(k,k) \ ,
\label{chi = ga}
\end{equation}
where $n_\pi$ is the norm of the pion vertex, which is defined with
eq.(\ref{normalizing}). This can also be checked by the
relation,
\begin{equation}
V {\cal S}(k) B(k) \gamma_5 {\cal S}(k) V = V { B(k) \over A(k)^2k^2 - B(k)^2} V \gamma_5 
\end{equation}
which explicitly verifies that the integrand of the Bethe Salpeter equation 
for the pion vertex, in the limit vanishing current quark mass $m$ and 
vertex momentum $P^\mu$, is identical to the integrand of the
Schwinger Dyson equation for the scalar component $B$ of the
dressed quark propagator.  It is important to remark that outside
this limit eq (\ref{chi = ga}) does not hold, but the difference between
the pion vertex and the axial vertex only starts to contribute at first
order in the expansion in $m$ and $P^\mu$,
\begin{equation}
\chi_{_P}(k) = 
{1 \over 2 \, i \, n_\pi } {\Gamma_{\hspace{-.08cm}A}}_P(k)+
o(P^\mu, m) 
\label{chi neq ga}
\end{equation}

\par
Substituting the spectral decomposition 
\cite{Pagels} of the ladder (\ref{ladder pole}), 
eq. (\ref{second}) implies that,
\begin{eqnarray}
{\Gamma_{\hspace{-.08cm}A}}_{P} 
& = & 
\chi_{P} { i \over P^2 -M_\pi^2 }  
tr\{ \chi_{_{-P}}
(S \gamma_{\hspace{-.08cm}A} S)_{P} \} 
  \ ,
\nonumber \\
\simeq
&&{1 \over 2 i\, n_\pi}\Gamma_{\hspace{-.08cm}A}{ i \over P^2 -M_\pi^2 }  
tr\{ (S \chi S)_{_{-P}}
{\gamma_{\hspace{-.08cm}A}}_{P} \}
 \ ,
\end{eqnarray}
where only the first non-vanishing terms in the expansion in 
$P^\mu$ and in $M_\pi$ is retained. In particular the pion vertex
in the left hand side was simplified with eq. (\ref{chi = ga}).
The leading result is,
\begin{equation}
tr\{ (S \chi S)_{_{-P}}
{\gamma_{\hspace{-.08cm}A}}_{_{P}} \}
= 2 \, n_\pi ( P^2 -M_\pi^2 ) \ .
\label{p2-m2}
\end{equation}
It is also convenient to extend this result to the case
where different external momenta are involved in 
$(S \chi S)$ and in $ \gamma_{\hspace{-.08cm}A} $.
A detailed momentum analysis of eqs (\ref{decay},\ref{bare},\ref{p2-m2})
shows that the norm $ n_\pi = i\, f_\pi / \sqrt{2}$ and 
produces 
\cite{Maris}
the desired trace,
\begin{equation}
tr\{ (S \chi S)_{_{P_1}}
{\gamma_{\hspace{-.08cm}A}}_{P_2} \}
=
 -2 \, n_\pi ( P_1 \cdot P_2 +M_\pi^2 ) \ .
\label{condensed}
\end{equation}
Using eqs. (\ref{condensed},\ref{dressed quark},\ref{bare}) when the momenta vanish, 
an important particular result is derived,
\begin{equation}
2 \, m \, tr\{ S \}
=
 f_\pi^2 \, M_\pi^2 \ ,
\label{GMOR}
\end{equation}
where $-tr\{ S \}$ is the quark condensate $\langle \bar \psi \psi\rangle$. 
Eq. (\ref{GMOR}) is the Gell-Mann Oakes and Renner relation
\cite{Gell-Mann}. In the chiral limit the quark condensate $\langle \bar \psi \psi\rangle$
and the pion decay constant $f_\pi$ remain constant. The Gell-Mann Oakes and Renner relation
shows that the pion mass $M_\pi$ is proportional to $\sqrt{m}$. 
The Gell-Mann Oakes and Renner relation can also be extended
\cite{Ivanov,Maris2,Roberts3} 
for arbitrarily large current-quark masses, and this
provides for instance an intuitive understanding of modern lattice simulations.

\section{The Goldberger-Treiman Relation} 

\par 
The Goldberger-Treiman relation provides a convenient
exercise to re-sum the series of planar diagrams,
to describe the intermediate virtual meson exchange 
with the ladder, and to check that the pion vertex 
can be used outside the mass shell.

\par
The weak decay of the neutron,
$n \rightarrow p + e^- + \bar \nu_e $
is computed with Feynman diagrams that include
the four Fermi coupling of two quark legs with the
electron and the neutrino leg. The bare loop of 
Fig. \ref{Goldberger Treiman} (a) does not correctly account
for the strong interaction. A complete set
of insertions of the quark-quark potential must be used.
However three classes of insertions are included from the onset.
The vertex $V$ of the quark-quark potential 
is assumed to be already renormalized and should not be further dressed. 
For instance renormalizing diagrams for $V$ have not been used neither
in the mass gap equation (\ref{mass gap}) nor in the ladder series (\ref{ladder}).
The propagator $S$ is also dressed, it is a solution of the mass gap equation
(\ref{mass gap}).
Moreover the Bethe Salpeter vertices $\chi$ of the proton and neutron
are already dressed, they
are solutions of the three body Bethe-Salpeter equation. 
However the four Fermi 
coupling is not dressed from the onset and it remains to be dressed.
A detailed inspection shows that the full series of planar diagrams 
can be re-summed in a ladder series which dresses 
\cite{Bloch,Maris3,Ji}
the Fermi coupling,
and in interactions in the remaining diquark of the nucleon.
The ladder series is represented by a box in 
Fig. \ref{Goldberger Treiman} (b).
The interactions in the remaining diquark of the nucleon are 
represented by an empty circle in Fig.  \ref{Goldberger Treiman} (b), 
however they will not affect the results of this paper.

\par
When the ladder in Fig. \ref{Goldberger Treiman} (b) is replaced by the
spectral decomposition of eq. (\ref{ladder pole}), the dressed
Feynman loop (b) factorizes in the product of the coupling of a nucleon to a pion
$ N \rightarrow P+\pi^-$,
of the pion propagator, and of the electroweak decay of the pion
$ \pi^- \rightarrow e^-+\nu_e$
\cite{Scadron}. 
The axial part of the electroweak decay can be measured, 
and I just have to compute the coupling of $P^\mu_1 \gamma_\mu \gamma_5$
to a quark line of the nucleon. This is included in the bare axial vertex
$\gamma_{\hspace{-.08cm}A}$ and  it is convenient to compute,
\FL
\begin{equation}
%
\begin{picture}(85,60)(0,0)
\put(0,0){
%
%
\put(20,40){
\begin{picture}(90,20)(0,0)
\put(12,15){$\gamma_{\hspace{-.08cm}A}$}
\put(18,8){$\bullet$}
\end{picture}}
\put(20,40){
\begin{picture}(40,20)(0,0)
\put(20,-5){\oval(10,29)[t]}
\put(10,-15){\framebox(20,10){}}
\put(25,0){\vector(0,1){5}}
\put(15,5){\vector(0,-1){5}}
\put(-10,0){$P_1$}
\put(-15,-5){\vector(0,1){15}}
\end{picture}}
\put(20,10){
\begin{picture}(40,20)(0,0)
\put(40,10){\line(-3,1){15}}
\put(15,15){\line(-3,-1){15}}
\put(40,10){\vector(-3,1){5}}
\put(15,15){\vector(-3,-1){13}}
\put(-5,12){$u$}
\put(-22,0){$P$}
\put(40,12){$d$}
\put(53,0){$N$}
\end{picture}}
\put(20,0){
\begin{picture}(40,20)(0,0)
\put(25,10){\line(1,0){25}}
\put(15,10){\line(-1,0){25}}
\put(15,10){\vector(-1,0){15}}
\put(50,10){\vector(-1,0){15}}
\end{picture}}
\put(25,-10){
\begin{picture}(40,20)(0,0)
\put(20,10){\line(1,0){15}}
\put(10,10){\line(-1,0){15}}
\put(10,10){\vector(-1,0){15}}
\put(35,10){\vector(-1,0){5}}
\end{picture}}
\put(20,-10){
\begin{picture}(40,20)(0,0)
\put(20,15){\circle{14}}
\end{picture}}
\put(20,-10){
\begin{picture}(40,20)(0,0)
\put(0,20){\oval(20,20)[l]}
\put(40,20){\oval(20,20)[r]}
\put(-12.5,18){$\bullet$}
\put(47.5,18){$\bullet$}
\end{picture}}
}
\end{picture}
%
=
%
%
\begin{picture}(55,35)(0,0)
\put(0,0){
%
%
\put(20,10){
\begin{picture}(40,20)(0,0)
\put(10,10){\line(-1,0){15}}
\put(10,10){\vector(-1,0){16}}
\put(10,10){\vector(-1,0){2}}
\put(-13,12){$u$}
\put(-24,0){$P$}
\put(15,12){$d$}
\put(23,0){$N$}
\end{picture}}
\put(20,10){
\begin{picture}(40,20)(0,0)
\put(-3,18){$\chi_\pi$}
\put(1,8){$\bullet$}
\end{picture}}
\put(20,0){
\begin{picture}(40,20)(0,0)
\put(20,10){\line(-1,0){12}}
\put(20,10){\vector(-1,0){9}}
\put(-2,10){\line(-1,0){13}}
\put(-2,10){\vector(-1,0){10}}
\end{picture}}
\put(20,-10){
\begin{picture}(40,20)(0,0)
\put(10,10){\line(-1,0){2}}
\put(13,10.5){\vector(-4,-1){1}}
\put(-2,10){\line(-1,0){3}}
\put(-11,12){\vector(-4,1){1}}
\end{picture}}
\put(20,-10){
\begin{picture}(40,20)(0,0)
\put(3,15){\circle{14}}
\end{picture}}
\put(20,-10){
\begin{picture}(40,20)(0,0)
\put(10,20){\oval(20,20)[r]}
\put(-5,20){\oval(20,20)[l]}
\put(-17.5,18){$\bullet$}
\put(17.5,18){$\bullet$}
\end{picture}}
}
\end{picture}
%
{ i \over P^2_1-M_\pi^2}
%
%
\begin{picture}(15,40)(0,0)
\put(0,0){
%
\put(0,10){
\begin{picture}(40,20)(0,0)
\put(5,10){\oval(10,20)}
\put(3,-3){$\bullet$}
\put(3,17){$\bullet$}
\put(0,10){\vector(0,-1){2}}
\put(10,10){\vector(0,1){2}}
\put(1,25){$\gamma_{\hspace{-.08cm}A}$}
\put(1,-8){$\chi_\pi$}
\end{picture}}
}
\end{picture}
%
\ .
\label{gt gammaa}
\end{equation}
Summing the contribution of the three quarks internal to the
proton and the neutron, the left hand side of eq. (\ref{gt gammaa}) 
is identical to $\sqrt{2} M_n g_A / i$.
This is defined in Nuclear Physics from the Dirac equation for the nucleon
which is considered as a Dirac fermion. The vertex $\gamma_A$ can be rewritten as
$( \not \hspace{-.05cm} k_P \gamma_5 + \gamma_5 \not \hspace{-.05cm} k_N -2m \gamma_5)/i$, 
where $k_P$ is the momentum of the quark that flows into the proton and 
$k_N$ is the momentum of the quark that comes from the neutron.
Summing the contribution of the three quarks to this amplitude,
and interpreting the nucleon as a Dirac particle,
the left hand side of eq. (\ref{gt gammaa}) is identical to the matrix element of 
$( \not \hspace{-.05cm} P_P \gamma_5 + \gamma_5 \not \hspace{-.05cm} P_N -6m \gamma_5)/i$.
Continuing to interpret the nucleon as a Dirac Particle the proton 
and nucleon slashed momenta can be replaced respectively by $M_N$ and $M_P$. 
The current quark mass $m$, and the mass difference
$M_N-M_P$ are both of the MeV order and negligible when compared with the 
nucleon mass.
The computation of the left hand side is completed with the matrix element of 
$\gamma_5$ in the Dirac nucleon which is $g_A$ except
for a possible phase.
In what concerns the right hand side of  eq. (\ref{gt gammaa}), the
sum in the three internal quark lines produces the coupling of the
pion to a nucleon. We do not need to concern with the  
interactions in the remaining diquark because they also dress
the pion coupling to the nucleon. 
Excluding phases eq. (\ref{gt gammaa}) is then,
\begin{equation}
2 M_n  g_A = \sqrt{2} 
g_{\pi n n} { 1 \over P^2_1-M_\pi^2 }\, \sqrt{2} \,f_\pi(P^2_1-M_\pi^2) ,
\label{for GT}
\end{equation}
where $P^\mu_1$ is the momentum that flows in the pseudoscalar ladder,
$-i \sqrt{2} g_{\pi n n}$ is the coupling of the
pion to the nucleon (the $\sqrt{2}$ is a flavor factor), and 
the pion decay constant $f_\pi$ is defined with the traces (\ref{decay}) and 
(\ref{condensed}).

\par
Eq. (\ref{for GT}) relates the nucleon decay with the pion decay constant,
the famous Goldberger Treiman relation,
\cite{Goldberger}, 
\begin{equation}
M_n \, g_A= g_{n \pi n} \, f_\pi  \ ,
\label{ GT relation}
\end{equation} 
which is correct except for a small discrepancy of 6\%
\cite{Pagels2,Ishii}.
This experimental verification suggests that
the planar series of diagrams is acceptable,
that the pseudoscalar ladder and the pion vertex 
can be used outside the mass shell, and that 
the expansion in $P^\mu$ and $M_\pi$ is convergent.

\section{The Adler Self-Consistency Zero} 

\par
Adler showed 
\cite{Adler}
that in the chiral limit of
$m=0$, pions of vanishing momentum decouple 
from other mesons on the mass shell. 
In this limit, see eq. (\ref{chi = ga}) the pion vertex is
proportional to the dressed axial vertex,
$\chi_\pi \ \alpha \ \Gamma_{\hspace{-.08cm}A}$. 
Therefore I simply have to show that $\Gamma_{\hspace{-.08cm}A}$
decouples from loops with meson vertices.
This decoupling is straightforward in a three meson coupling,
\FL
\begin{eqnarray}
\begin{picture}(75,40)(0,0)
\put(0,8){
\put(20,0){\begin{picture}(40,40)(0,0)
\put(0,0){\line(1,0){30}}
\put(0,0){\vector(1,0){15}}
\put(30,0){\line(-1,1){15}}
\put(30,0){\vector(-1,1){8}}
\put(15,15){\line(-1,-1){15}}
\put(15,15){\vector(-1,-1){8}}
\put(-2,-2){$\bullet$}
\put(-20,-2){${\chi_3}_{P_3}$}
\put(28,-2){$\bullet$}
\put(35,-2){${\chi_2}_{P_2}$}
\put(13,13){$\bullet$}
\put(15,20){${\Gamma_{\hspace{-.08cm}A}}_{P_1}$}
\end{picture}}
}
\end{picture}
&=&
{P_3^2-M_3^2 \over i {\cal I}_3 } \ {P_2^2-M_2^2 \over i {\cal I}_2 } \ \times
\nonumber \\
&&
\begin{picture}(135,40)(0,0)
%
\put(23,10){
\begin{picture}(40,20)(0,0)
\put(0,15){\line(1,0){5}}
\put(0,5){\vector(1,0){10}}
\put(15,15){\vector(-1,0){10}}
\put(15,5){\line(-1,0){5}}
\put(15,0){\framebox(10,20){}}
\put(25,15){\line(1,0){5}}
\put(25,5){\vector(1,0){10}}
\put(40,15){\vector(-1,0){10}}
\put(40,5){\line(-1,0){5}}
\end{picture}}
\put(63,10){
\begin{picture}(40,20)(0,0)
\put(0,15){\line(1,0){5}}
\put(0,5){\vector(1,0){10}}
\put(15,15){\vector(-1,0){10}}
\put(15,5){\line(-1,0){5}}
\put(15,0){\framebox(10,20){}}
\put(25,15){\line(1,0){5}}
\put(25,5){\vector(1,0){10}}
\put(40,15){\vector(-1,0){10}}
\put(40,5){\line(-1,0){5}}
\end{picture}}
\put(-22,10){
\begin{picture}(40,20)(0,0)
\put(45,10){\oval(10,10)[l]}
\put(37,8){$\bullet$}
\put(20,7){${\chi_3}_{P_3}$}
\end{picture}}
\put(63,10){
\begin{picture}(40,20)(0,0)
\put(40,10){\oval(10,10)[r]}
\put(43,8){$\bullet$}
\put(49,10){$ {\chi_2}_{P_2}$}
\end{picture}}
\put(23,10){
\begin{picture}(40,20)(0,0)
\put(40,2){$\bullet$}
\put(40,13){$\bullet$}
\put(30,-6){$ S^{-1}$}
\put(30,21){$ {\Gamma_{\hspace{-.08cm}A}}_{P_1}$}
\end{picture}}
\end{picture}
\nonumber \\
&=&
{P_3^2-M_3^2 \over i {\cal I}_3 } \ \
\begin{picture}(90,20)(0,0)
%
\put(43,0){
\begin{picture}(40,20)(0,0)
\put(0,15){\line(1,0){5}}
\put(0,5){\vector(1,0){10}}
\put(15,15){\vector(-1,0){10}}
\put(15,5){\line(-1,0){5}}
\end{picture}}
\put(18,0){
\begin{picture}(40,20)(0,0)
\put(25,10){\oval(10,10)[l]}
\put(17,8){$\bullet$}
\put(-20,7){${\chi_3}_{P_3}{\gamma_5}_{P_1}$}
\end{picture}}
\put(58,0){
\begin{picture}(40,20)(0,0)
\put(0,10){\oval(10,10)[r]}
\put(3,8){$\bullet$}
\put(9,10){$ {\chi_2}_{P_2}$}
\end{picture}}
\end{picture}
\nonumber \\
&&+
{P_2^2-M_2^2 \over i {\cal I}_2 } \ \
\begin{picture}(90,20)(0,0)
%
\put(25,0){
\begin{picture}(40,20)(0,0)
\put(0,15){\line(1,0){5}}
\put(0,5){\vector(1,0){10}}
\put(15,15){\vector(-1,0){10}}
\put(15,5){\line(-1,0){5}}
\end{picture}}
\put(0,0){
\begin{picture}(40,20)(0,0)
\put(25,10){\oval(10,10)[l]}
\put(17,8){$\bullet$}
\put(-2,7){${\chi_3}_{P_3}$}
\end{picture}}
\put(40,0){
\begin{picture}(40,20)(0,0)
\put(0,10){\oval(10,10)[r]}
\put(3,8){$\bullet$}
\put(9,10){${\gamma_5}_{P_1} {\chi_2}_{P_2}$}
\end{picture}}
\end{picture}
\nonumber \\
&=& 0
\label{3 meson}
\end{eqnarray}
and this vanishes when the mesons of vertex $\chi_2$
and $\chi_3$ are on the mass shell. To get this result
I used eqs. (\ref{pole pion}) and (\ref{crucial}).
In the three meson coupling of eq. (\ref{3 meson}) the
Feynman loop is empty. Any planar insertion
of the quark-quark interaction would produce double counting,
because the vertices are already dressed.

\par
However the three meson coupling is not the best one
to find the direct evidence of an Adler zero.
Because the mesons of vertex  $\chi_2$ and 
$\chi_3$ couple to a pion, either the coupling is derivative,
or the mesons $1$ and $2$ have opposite parity.
In the case of a derivative coupling the vanishing result is trivial,
and it is not a PCAC result. 
In the case of opposite parity, and 
because chiral symmetry is spontaneously broken, the mesons 
$1$ and $2$ are not expected to have the same mass. Therefore either
$1$ and $2$ are not both on the mass shell, or the pion momentum
in not vanishing, and eq. (\ref{3 meson}) does not apply.

\par
The four meson coupling is more interesting than the three meson one.  
If two pions are coupled to two identical mesons, then all four mesons
can be on the mass shell. 
In the coupling of four mesons, the planar diagrams must be
included, and they can be re-summed in two different intermediate
ladder exchanges,
%
\begin{equation}
\begin{picture}(55,55)(0,0)
\put(0,-8){
\put(16,13){\begin{picture}(20,40)(0,0)
\put(5,0){\line(0,1){5}}
\put(5,15){\vector(0,-1){10}}
\put(15,15){\line(0,-1){5}}
\put(15,0){\vector(0,1){10}}
\put(0,15){\framebox(20,10){}}
\put(5,25){\line(0,1){5}}
\put(5,40){\vector(0,-1){10}}
\put(15,40){\line(0,-1){5}}
\put(15,25){\vector(0,1){10}}
\end{picture}}
\put(16,8){\begin{picture}(20,40)(0,0)
\put(10,40){\oval(10,26)[t]}
\put(2,42){$\bullet$}
\put(-13,48){$\chi_{P_4}$}
\put(13,42){$\bullet$}
\put(18,48){${\Gamma_{\hspace{-.08cm}A}}_{P_1}$}
\put(10,53){\vector(-1,0){2}}
\end{picture}}
\put(16,-27){\begin{picture}(20,40)(0,0)
\put(10,45){\oval(10,26)[b]}
\put(2,38){$\bullet$}
\put(-15,34){$\chi_{P_3}$}
\put(13,38){$\bullet$}
\put(18,34){$\chi_{P_2}$}
\put(10,32){\vector(1,0){2}}
\end{picture}}
}
\end{picture}
+
\begin{picture}(80,30)(0,0)
\put(0,-8){
\put(20,10){\begin{picture}(40,20)(0,0)
\put(0,15){\line(1,0){5}}
\put(0,5){\vector(1,0){10}}
\put(15,15){\vector(-1,0){10}}
\put(15,5){\line(-1,0){5}}
\put(15,0){\framebox(10,20){}}
\put(25,15){\line(1,0){5}}
\put(25,5){\vector(1,0){10}}
\put(40,15){\vector(-1,0){10}}
\put(40,5){\line(-1,0){5}}
\end{picture}}
\put(20,10){\begin{picture}(40,20)(0,0)
\put(40,10){\oval(30,10)[r]}
\put(42,2){$\bullet$}
\put(38,-5){$\chi_{P_2}$}
\put(42,13){$\bullet$}
\put(38,21){${\Gamma_{\hspace{-.08cm}A}}_{P_1}$}
\put(55,10){\vector(0,1){2}}
\end{picture}}
\put(-25,10){\begin{picture}(40,20)(0,0)
\put(45,10){\oval(30,10)[l]}
\put(38,2){$\bullet$}
\put(34,-5){$\chi_{P_3}$}
\put(38,13){$\bullet$}
\put(34,21){$\chi_{P_4}$}
\put(30,10){\vector(0,-1){2}}
\end{picture}}
}
\end{picture}
-
\begin{picture}(55,30)(0,0)
\put(0,-8){
\put(20,10){\begin{picture}(40,20)(0,0)
\put(0,15){\line(1,0){5}}
\put(0,5){\vector(1,0){10}}
\put(15,15){\vector(-1,0){10}}
\put(15,5){\line(-1,0){5}}
\end{picture}}
\put(-5,10){\begin{picture}(40,20)(0,0)
\put(40,10){\oval(30,10)[r]}
\put(42,2){$\bullet$}
\put(38,-5){$\chi_{P_2}$}
\put(42,13){$\bullet$}
\put(38,21){${\Gamma_{\hspace{-.08cm}A}}_{P_1}$}
\put(55,10){\vector(0,1){2}}
\end{picture}}
\put(-25,10){\begin{picture}(40,20)(0,0)
\put(45,10){\oval(30,10)[l]}
\put(38,2){$\bullet$}
\put(34,-5){${\Gamma_{\hspace{-.08cm}A}}_{P_3}$}
\put(38,13){$\bullet$}
\put(34,21){$\chi_{P_4}$}
\put(30,10){\vector(0,-1){2}}
\end{picture}}
}
\end{picture}
\ ,
\label{4 meson}
\end{equation}
where the empty box is subtracted to cure double counting. 
I again follow the prescription defined in Section IV of
excluding diagrams which would dress the quark-quark potential
vertex $V$, the quark propagator $\cal S$ or the meson vertex
$\chi_{P_i}$.
The intermediate ladders include both a direct contact term, and the pole
corresponding to meson exchange. 
There is also evidence 
that the hadron-hadron 
coupled channel equations should include one meson
exchange, 
both in the sigma model 
\cite{miracle},
in the Nambu and Jona-Lasinio model 
\cite{Veronique}, 
in the constituent Quark Models
\cite{pi-pi,Goncalo},
and in an Euclidean Quark Model 
\cite{pi-pi}.
In microscopic calculations the Feynman loop of
a four meson coupling, which dominates for instance
for $\pi-\pi$ scattering, must therefore 
include inside the box a vertical scalar ladder and a 
horizontal scalar ladder \cite{pi-pi}.

\par
The main step to get the zero consists in decreasing the number of vertices
using again eqs. (\ref{pole pion}) and 
(\ref{crucial}). For instance I find that the second diagram
of eq(\ref{4 meson}), 
\begin{equation}
\begin{picture}(80,30)(0,0)
\put(0,-8){
\put(20,10){\begin{picture}(40,20)(0,0)
\put(0,15){\line(1,0){5}}
\put(0,5){\vector(1,0){10}}
\put(15,15){\vector(-1,0){10}}
\put(15,5){\line(-1,0){5}}
\put(15,0){\framebox(10,20){}}
\put(25,15){\line(1,0){5}}
\put(25,5){\vector(1,0){10}}
\put(40,15){\vector(-1,0){10}}
\put(40,5){\line(-1,0){5}}
\end{picture}}
\put(20,10){\begin{picture}(40,20)(0,0)
\put(40,10){\oval(30,10)[r]}
\put(42,2){$\bullet$}
\put(38,-5){$\chi_{P_2}$}
\put(42,13){$\bullet$}
\put(38,21){${\Gamma_{\hspace{-.08cm}A}}_{P_1}$}
\put(55,10){\vector(0,1){2}}
\end{picture}}
\put(-25,10){\begin{picture}(40,20)(0,0)
\put(45,10){\oval(30,10)[l]}
\put(38,2){$\bullet$}
\put(34,-5){$\chi_{P_3}$}
\put(38,13){$\bullet$}
\put(34,21){$\chi_{P_4}$}
\put(30,10){\vector(0,-1){2}}
\end{picture}}
}
\end{picture}
\end{equation}
is identical to,
\begin{eqnarray}
&&
{P_2^2-M_2^2 \over i {\cal I}_2 }
\begin{picture}(135,30)(0,0)
%
\put(23,0){
\begin{picture}(40,20)(0,0)
\put(0,15){\line(1,0){5}}
\put(0,5){\vector(1,0){10}}
\put(15,15){\vector(-1,0){10}}
\put(15,5){\line(-1,0){5}}
\put(15,0){\framebox(10,20){}}
\put(25,15){\line(1,0){5}}
\put(25,5){\vector(1,0){10}}
\put(40,15){\vector(-1,0){10}}
\put(40,5){\line(-1,0){5}}
\end{picture}}
\put(63,0){
\begin{picture}(40,20)(0,0)
\put(0,15){\line(1,0){5}}
\put(0,5){\vector(1,0){10}}
\put(15,15){\vector(-1,0){10}}
\put(15,5){\line(-1,0){5}}
\put(15,0){\framebox(10,20){}}
\put(25,15){\line(1,0){5}}
\put(25,5){\vector(1,0){10}}
\put(40,15){\vector(-1,0){10}}
\put(40,5){\line(-1,0){5}}
\end{picture}}
\put(-17,0)
{\begin{picture}(40,20)(0,0)
\put(45,10){\oval(30,10)[l]}
\put(38,2){$\bullet$}
\put(34,-5){$\chi_{P_3}$}
\put(38,13){$\bullet$}
\put(34,21){$\chi_{P_4}$}
\put(30,10){\vector(0,-1){2}}
\end{picture}}
\put(63,0){
\begin{picture}(40,20)(0,0)
\put(40,10){\oval(10,10)[r]}
\put(43,8){$\bullet$}
\put(49,10){$ {\chi}_{P_2}$}
\end{picture}}
\put(23,0){
\begin{picture}(40,20)(0,0)
\put(40,2){$\bullet$}
\put(40,13){$\bullet$}
\put(30,-6){$ S^{-1}$}
\put(30,21){$ {\Gamma_{\hspace{-.08cm}A}}_{P_1}$}
\end{picture}}
\end{picture}
=
\\ \nonumber 
&&
\begin{picture}(70,40)(0,0)
%
\put(18,0){
\begin{picture}(40,20)(0,0)
\put(0,15){\line(1,0){5}}
\put(0,5){\vector(1,0){10}}
\put(15,15){\vector(-1,0){10}}
\put(15,5){\line(-1,0){5}}
\end{picture}}
\put(-22,0)
{\begin{picture}(40,20)(0,0)
\put(45,10){\oval(30,10)[l]}
\put(38,2){$\bullet$}
\put(34,-5){$\chi_{P_3}$}
\put(38,13){$\bullet$}
\put(24,22){$\chi_{P_4}{\gamma_5}_{P_1}$}
\put(30,10){\vector(0,-1){2}}
\end{picture}}
\put(33,0){
\begin{picture}(40,20)(0,0)
\put(0,10){\oval(10,10)[r]}
\put(3,8){$\bullet$}
\put(9,10){$ {\chi}_{P_2}$}
\end{picture}}
\end{picture}
+
{P_2^2-M_2^2 \over i {\cal I}_2 } \ \
\begin{picture}(110,40)(0,0)
%
\put(15,0){
\begin{picture}(40,20)(0,0)
\put(0,15){\line(1,0){5}}
\put(0,5){\vector(1,0){10}}
\put(15,15){\vector(-1,0){10}}
\put(15,5){\line(-1,0){5}}
\put(15,0){\framebox(10,20){}}
\put(25,15){\line(1,0){5}}
\put(25,5){\vector(1,0){10}}
\put(40,15){\vector(-1,0){10}}
\put(40,5){\line(-1,0){5}}
\end{picture}}
\put(-25,0)
{\begin{picture}(40,20)(0,0)
\put(45,10){\oval(30,10)[l]}
\put(38,2){$\bullet$}
\put(34,-5){$\chi_{P_3}$}
\put(38,13){$\bullet$}
\put(34,21){$\chi_{P_4}$}
\put(30,10){\vector(0,-1){2}}
\end{picture}}
\put(55,0){
\begin{picture}(40,20)(0,0)
\put(0,10){\oval(10,10)[r]}
\put(3,8){$\bullet$}
\put(9,10){${\gamma_5}_{P_1} {\chi}_{P_2}$}
\end{picture}}
\end{picture}
\end{eqnarray}
The first diagram of eq. (\ref{4 meson}) is computed in
the same way.
The crucial step consists in realizing that in the sum of the 
three diagrams of eq. (\ref{4 meson}),
the empty loop, without intermediate
ladders, exactly cancels due to the eq. (\ref{axialWI}).
The sum of the three diagrams of eq. (\ref{4 meson})
is exactly equal to,
\begin{eqnarray}
{P_4^2-M_4^2 \over i {\cal I}_4 } \ \
\begin{picture}(110,30)(0,0)
%
\put(15,0){
\begin{picture}(40,20)(0,0)
\put(0,15){\line(1,0){5}}
\put(0,5){\vector(1,0){10}}
\put(15,15){\vector(-1,0){10}}
\put(15,5){\line(-1,0){5}}
\put(15,0){\framebox(10,20){}}
\put(25,15){\line(1,0){5}}
\put(25,5){\vector(1,0){10}}
\put(40,15){\vector(-1,0){10}}
\put(40,5){\line(-1,0){5}}
\end{picture}}
\put(-25,0)
{\begin{picture}(40,20)(0,0)
\put(45,10){\oval(30,10)[l]}
\put(38,2){$\bullet$}
\put(34,-5){$\chi_{P_2}$}
\put(38,13){$\bullet$}
\put(34,21){$\chi_{P_3}$}
\put(30,10){\vector(0,-1){2}}
\end{picture}}
\put(55,0){
\begin{picture}(40,20)(0,0)
\put(0,10){\oval(10,10)[r]}
\put(3,8){$\bullet$}
\put(9,10){$ {\chi}_{P_4}{\gamma_5}_{P_1}$}
\end{picture}}
\end{picture}
+&&
\nonumber \\
{P_2^2-M_2^2 \over i {\cal I}_2 } \ \
\begin{picture}(110,30)(0,0)
%
\put(15,0){
\begin{picture}(40,20)(0,0)
\put(0,15){\line(1,0){5}}
\put(0,5){\vector(1,0){10}}
\put(15,15){\vector(-1,0){10}}
\put(15,5){\line(-1,0){5}}
\put(15,0){\framebox(10,20){}}
\put(25,15){\line(1,0){5}}
\put(25,5){\vector(1,0){10}}
\put(40,15){\vector(-1,0){10}}
\put(40,5){\line(-1,0){5}}
\end{picture}}
\put(-25,0)
{\begin{picture}(40,20)(0,0)
\put(45,10){\oval(30,10)[l]}
\put(38,2){$\bullet$}
\put(34,-5){$\chi_{P_3}$}
\put(38,13){$\bullet$}
\put(34,21){$\chi_{P_4}$}
\put(30,10){\vector(0,-1){2}}
\end{picture}}
\put(55,0){
\begin{picture}(40,20)(0,0)
\put(0,10){\oval(10,10)[r]}
\put(3,8){$\bullet$}
\put(9,10){${\gamma_5}_{P_1}{\chi}_{P_2}$}
\end{picture}}
\end{picture}
&&=0
\label{result 4 meson}
\end{eqnarray}
and this vanishes when the mesons of vertex $\chi_2$ and
$\chi_4$ are on the mass shell. Although poles occur
in the remaining intermediate ladder in eq. (\ref{result 4 meson}),
they are not expected to reside, say at $P_2^2=M_2^2$, 
because $\chi_2 {\gamma_5}_{P_1}$ has the opposite parity of $\chi_2$,
and because chiral symmetry is spontaneously broken.

\par
The same method can be used to show that the pion
{\em decouples from any number of mesons } on the mass shell. 
This constitutes a Ward identity for the meson couplings.
The Quark Model complies with the Adler self consistency zero.

\section{The Weinberg Theorem} 

\par
The four pion coupling, which dominates $\pi-\pi$ low energy scattering
is the ideal process where the Adler self consistency zero applies.
To find a non-vanishing contribution, the Feynman loop which extends
eq. (\ref{4 meson}) 
must be expanded up to first order in  $P_i\cdot P_j$ and in $M_\pi^2$. 
The result is a beautiful algebraic expression which was first derived 
by Weinberg. After the original work of Weinberg
\cite{Weinberg}, 
the theorem was also derived 
with Ward identities for the pion fields 
\cite{Leutwyler} 
and  with a functional integration of quarks
\cite{Roberts}.
I now prove that the $\pi-\pi$ scattering theorem 
of Weinberg 
\cite{Weinberg} 
applies to Quark Models with chiral 
invariant quark-quark interactions,
completing in full detail an 
analytical proof which was recently outlined 
in \cite{pi-pi,Goncalo}.

\par
The most technical task of this paper consists in computing 
{\em independently of the Quark Model},
and up to order $M_\pi^2$ and $P_i \, P_j$
in the $\pi$ mass and momenta,
the Feynman loop,
%
%
\begin{equation}
\begin{picture}(55,55)(0,0)
\put(0,-8){
\put(16,13){\begin{picture}(20,40)(0,0)
\put(5,0){\line(0,1){5}}
\put(5,15){\vector(0,-1){10}}
\put(15,15){\line(0,-1){5}}
\put(15,0){\vector(0,1){10}}
\put(0,15){\framebox(20,10){}}
\put(5,25){\line(0,1){5}}
\put(5,40){\vector(0,-1){10}}
\put(15,40){\line(0,-1){5}}
\put(15,25){\vector(0,1){10}}
\end{picture}}
\put(16,8){\begin{picture}(20,40)(0,0)
\put(10,40){\oval(10,26)[t]}
\put(2,42){$\bullet$}
\put(-13,48){$\chi_{P_4}$}
\put(13,42){$\bullet$}
\put(18,48){$\chi_{P_1}$}
\put(10,53){\vector(-1,0){2}}
\end{picture}}
\put(16,-27){\begin{picture}(20,40)(0,0)
\put(10,45){\oval(10,26)[b]}
\put(2,38){$\bullet$}
\put(-15,34){$\chi_{P_3}$}
\put(13,38){$\bullet$}
\put(18,34){$\chi_{P_2}$}
\put(10,32){\vector(1,0){2}}
\end{picture}}
}
\end{picture}
+
\begin{picture}(80,30)(0,0)
\put(0,-8){
\put(20,10){\begin{picture}(40,20)(0,0)
\put(0,15){\line(1,0){5}}
\put(0,5){\vector(1,0){10}}
\put(15,15){\vector(-1,0){10}}
\put(15,5){\line(-1,0){5}}
\put(15,0){\framebox(10,20){}}
\put(25,15){\line(1,0){5}}
\put(25,5){\vector(1,0){10}}
\put(40,15){\vector(-1,0){10}}
\put(40,5){\line(-1,0){5}}
\end{picture}}
\put(20,10){\begin{picture}(40,20)(0,0)
\put(40,10){\oval(30,10)[r]}
\put(42,2){$\bullet$}
\put(38,-5){$\chi_{P_2}$}
\put(42,13){$\bullet$}
\put(38,21){$\chi_{P_1}$}
\put(55,10){\vector(0,1){2}}
\end{picture}}
\put(-25,10){\begin{picture}(40,20)(0,0)
\put(45,10){\oval(30,10)[l]}
\put(38,2){$\bullet$}
\put(34,-5){$\chi_{P_3}$}
\put(38,13){$\bullet$}
\put(34,21){$\chi_{P_4}$}
\put(30,10){\vector(0,-1){2}}
\end{picture}}
}
\end{picture}
-
\begin{picture}(55,30)(0,0)
\put(0,-8){
\put(20,10){\begin{picture}(40,20)(0,0)
\put(0,15){\line(1,0){5}}
\put(0,5){\vector(1,0){10}}
\put(15,15){\vector(-1,0){10}}
\put(15,5){\line(-1,0){5}}
\end{picture}}
\put(-5,10){\begin{picture}(40,20)(0,0)
\put(40,10){\oval(30,10)[r]}
\put(42,2){$\bullet$}
\put(38,-5){$\chi_{P_2}$}
\put(42,13){$\bullet$}
\put(38,21){$\chi_{P_1}$}
\put(55,10){\vector(0,1){2}}
\end{picture}}
\put(-25,10){\begin{picture}(40,20)(0,0)
\put(45,10){\oval(30,10)[l]}
\put(38,2){$\bullet$}
\put(34,-5){$\chi_{P_3}$}
\put(38,13){$\bullet$}
\put(34,21){$\chi_{P_4}$}
\put(30,10){\vector(0,-1){2}}
\end{picture}}
}
\end{picture}
\ ,
\label{loop}
\end{equation}
where $\chi$ is
the Bethe Salpeter vertex of the pion. The
subindex $_{Pi}$ accounts an external momentum flowing into
the loop.

\par
To get the loop (\ref{loop}) up to order $P_i \cdot P_j$ and
$M_\pi^2$,
at most two full Bethe Salpeter vertices
$\chi$ are needed. The other two can be approximated by 
$\Gamma_{\hspace{-.08cm}A} / (2 \, i \, n_\pi)$, according
to eq. (\ref{chi neq ga}).
Expanding the four $\chi$, which are respectively 
equal to 
$ \Gamma_{\hspace{-.08cm}A}/ (2 \, i \, n_\pi ) 
+[ \chi- \Gamma_{\hspace{-.08cm}A}/ (2 \, i \, n_\pi )]$,  
up to second order in 
$ \chi-{ \Gamma_{\hspace{-.08cm}A} / (2 \, i \, n_\pi} ) $
and regrouping the sum, one finds that the amplitude
of eq.(\ref{loop}) is the sum of four classes
of terms.  Each class includes a sum of the possible 
cyclic permutations of the external momenta 
$P_1, \, P_2, \, P_3,$ and $P_4$.
The sum of the four classes is identical to, 
with factor 3 times $(2 \, i \, n_\pi)^{-4}$ ,
\begin{equation}
\begin{picture}(55,55)(0,8)
\put(0,0){
\put(16,13){\begin{picture}(20,40)(0,0)
\put(5,0){\line(0,1){5}}
\put(5,15){\vector(0,-1){10}}
\put(15,15){\line(0,-1){5}}
\put(15,0){\vector(0,1){10}}
\put(0,15){\framebox(20,10){}}
\put(5,25){\line(0,1){5}}
\put(5,40){\vector(0,-1){10}}
\put(15,40){\line(0,-1){5}}
\put(15,25){\vector(0,1){10}}
\end{picture}}
\put(16,8){\begin{picture}(20,40)(0,0)
\put(10,40){\oval(10,26)[t]}
\put(2,40){$\bullet$}
\put(-15,49){${\Gamma_{\hspace{-.08cm}A}}_{P_4}$}
\put(13,42){$\bullet$}
\put(18,48){${\Gamma_{\hspace{-.08cm}A}}_{P_1}$}
\put(10,53){\vector(-1,0){2}}
\end{picture}}
\put(16,-27){\begin{picture}(20,40)(0,0)
\put(10,45){\oval(10,26)[b]}
\put(2,38){$\bullet$}
\put(-15,34){${\Gamma_{\hspace{-.08cm}A}}_{P_3}$}
\put(13,38){$\bullet$}
\put(18,34){${\Gamma_{\hspace{-.08cm}A}}_{P_2}$}
\put(10,32){\vector(1,0){2}}
\end{picture}}
}
\end{picture}
+
\begin{picture}(80,30)(0,0)
\put(0,0){
\put(20,10){\begin{picture}(40,20)(0,0)
\put(0,15){\line(1,0){5}}
\put(0,5){\vector(1,0){10}}
\put(15,15){\vector(-1,0){10}}
\put(15,5){\line(-1,0){5}}
\put(15,0){\framebox(10,20){}}
\put(25,15){\line(1,0){5}}
\put(25,5){\vector(1,0){10}}
\put(40,15){\vector(-1,0){10}}
\put(40,5){\line(-1,0){5}}
\end{picture}}
\put(20,10){\begin{picture}(40,20)(0,0)
\put(40,10){\oval(30,10)[r]}
\put(42,2){$\bullet$}
\put(38,-5){${\Gamma_{\hspace{-.08cm}A}}_{P_2}$}
\put(42,13){$\bullet$}
\put(38,21){${\Gamma_{\hspace{-.08cm}A}}_{P_1}$}
\put(55,10){\vector(0,1){2}}
\end{picture}}
\put(-25,10){\begin{picture}(40,20)(0,0)
\put(45,10){\oval(30,10)[l]}
\put(38,2){$\bullet$}
\put(34,-5){${\Gamma_{\hspace{-.08cm}A}}_{P_3}$}
\put(38,13){$\bullet$}
\put(34,21){${\Gamma_{\hspace{-.08cm}A}}_{P_4}$}
\put(30,10){\vector(0,-1){2}}
\end{picture}}
}
\end{picture}
-
\begin{picture}(55,30)(0,0)
\put(0,0){
\put(20,10){\begin{picture}(40,20)(0,0)
\put(0,15){\line(1,0){5}}
\put(0,5){\vector(1,0){10}}
\put(15,15){\vector(-1,0){10}}
\put(15,5){\line(-1,0){5}}
\end{picture}}
\put(-5,10){\begin{picture}(40,20)(0,0)
\put(40,10){\oval(30,10)[r]}
\put(42,2){$\bullet$}
\put(38,-5){${\Gamma_{\hspace{-.08cm}A}}_{P_2}$}
\put(42,13){$\bullet$}
\put(38,21){${\Gamma_{\hspace{-.08cm}A}}_{P_1}$}
\put(55,10){\vector(0,1){2}}
\end{picture}}
\put(-25,10){\begin{picture}(40,20)(0,0)
\put(45,10){\oval(30,10)[l]}
\put(38,2){$\bullet$}
\put(34,-5){${\Gamma_{\hspace{-.08cm}A}}_{P_3}$}
\put(38,13){$\bullet$}
\put(34,21){${\Gamma_{\hspace{-.08cm}A}}_{P_4}$}
\put(30,10){\vector(0,-1){2}}
\end{picture}}
}
\end{picture}
\label{ciclic 1}
\end{equation}
minus , with factor $2$ times $(2 \, i \, n_\pi)^{-3}$ , (4 permutations)
%
%
\begin{equation}
\begin{picture}(55,55)(0,0)
\put(0,-8){
\put(16,13){\begin{picture}(20,40)(0,0)
\put(5,0){\line(0,1){5}}
\put(5,15){\vector(0,-1){10}}
\put(15,15){\line(0,-1){5}}
\put(15,0){\vector(0,1){10}}
\put(0,15){\framebox(20,10){}}
\put(5,25){\line(0,1){5}}
\put(5,40){\vector(0,-1){10}}
\put(15,40){\line(0,-1){5}}
\put(15,25){\vector(0,1){10}}
\end{picture}}
\put(16,8){\begin{picture}(20,40)(0,0)
\put(10,40){\oval(10,26)[t]}
\put(2,42){$\bullet$}
\put(-13,48){$\chi_{P_4}$}
\put(13,42){$\bullet$}
\put(18,48){${\Gamma_{\hspace{-.08cm}A}}_{P_1}$}
\put(10,53){\vector(-1,0){2}}
\end{picture}}
\put(16,-27){\begin{picture}(20,40)(0,0)
\put(10,45){\oval(10,26)[b]}
\put(2,38){$\bullet$}
\put(-15,34){${\Gamma_{\hspace{-.08cm}A}}_{P_3}$}
\put(13,38){$\bullet$}
\put(18,34){${\Gamma_{\hspace{-.08cm}A}}_{P_2}$}
\put(10,32){\vector(1,0){2}}
\end{picture}}
}
\end{picture}
+
\begin{picture}(80,30)(0,0)
\put(0,-8){
\put(20,10){\begin{picture}(40,20)(0,0)
\put(0,15){\line(1,0){5}}
\put(0,5){\vector(1,0){10}}
\put(15,15){\vector(-1,0){10}}
\put(15,5){\line(-1,0){5}}
\put(15,0){\framebox(10,20){}}
\put(25,15){\line(1,0){5}}
\put(25,5){\vector(1,0){10}}
\put(40,15){\vector(-1,0){10}}
\put(40,5){\line(-1,0){5}}
\end{picture}}
\put(20,10){\begin{picture}(40,20)(0,0)
\put(40,10){\oval(30,10)[r]}
\put(42,2){$\bullet$}
\put(38,-5){${\Gamma_{\hspace{-.08cm}A}}_{P_2}$}
\put(42,13){$\bullet$}
\put(38,21){${\Gamma_{\hspace{-.08cm}A}}_{P_1}$}
\put(55,10){\vector(0,1){2}}
\end{picture}}
\put(-25,10){\begin{picture}(40,20)(0,0)
\put(45,10){\oval(30,10)[l]}
\put(38,2){$\bullet$}
\put(34,-5){${\Gamma_{\hspace{-.08cm}A}}_{P_3}$}
\put(38,13){$\bullet$}
\put(34,21){$\chi_{P_4}$}
\put(30,10){\vector(0,-1){2}}
\end{picture}}
}
\end{picture}
-
\begin{picture}(55,30)(0,0)
\put(0,-8){
\put(20,10){\begin{picture}(40,20)(0,0)
\put(0,15){\line(1,0){5}}
\put(0,5){\vector(1,0){10}}
\put(15,15){\vector(-1,0){10}}
\put(15,5){\line(-1,0){5}}
\end{picture}}
\put(-5,10){\begin{picture}(40,20)(0,0)
\put(40,10){\oval(30,10)[r]}
\put(42,2){$\bullet$}
\put(38,-5){${\Gamma_{\hspace{-.08cm}A}}_{P_2}$}
\put(42,13){$\bullet$}
\put(38,21){${\Gamma_{\hspace{-.08cm}A}}_{P_1}$}
\put(55,10){\vector(0,1){2}}
\end{picture}}
\put(-25,10){\begin{picture}(40,20)(0,0)
\put(45,10){\oval(30,10)[l]}
\put(38,2){$\bullet$}
\put(34,-5){${\Gamma_{\hspace{-.08cm}A}}_{P_3}$}
\put(38,13){$\bullet$}
\put(34,21){$\chi_{P_4}$}
\put(30,10){\vector(0,-1){2}}
\end{picture}}
}
\end{picture}
\ ,
\label{ciclic 2}
\end{equation}
plus, with factor $(2 \, i \, n_\pi)^{-2}$ , (4 permutations)
%
%
\begin{equation}
\begin{picture}(55,55)(0,0)
\put(0,-8){
\put(16,13){\begin{picture}(20,40)(0,0)
\put(5,0){\line(0,1){5}}
\put(5,15){\vector(0,-1){10}}
\put(15,15){\line(0,-1){5}}
\put(15,0){\vector(0,1){10}}
\put(0,15){\framebox(20,10){}}
\put(5,25){\line(0,1){5}}
\put(5,40){\vector(0,-1){10}}
\put(15,40){\line(0,-1){5}}
\put(15,25){\vector(0,1){10}}
\end{picture}}
\put(16,8){\begin{picture}(20,40)(0,0)
\put(10,40){\oval(10,26)[t]}
\put(2,42){$\bullet$}
\put(-13,48){$\chi_{P_4}$}
\put(13,42){$\bullet$}
\put(18,48){${\Gamma_{\hspace{-.08cm}A}}_{P_1}$}
\put(10,53){\vector(-1,0){2}}
\end{picture}}
\put(16,-27){\begin{picture}(20,40)(0,0)
\put(10,45){\oval(10,26)[b]}
\put(2,38){$\bullet$}
\put(-15,34){$\chi_{P_3}$}
\put(13,38){$\bullet$}
\put(18,34){${\Gamma_{\hspace{-.08cm}A}}_{P_2}$}
\put(10,32){\vector(1,0){2}}
\end{picture}}
}
\end{picture}
+
\begin{picture}(80,30)(0,0)
\put(0,-8){
\put(20,10){\begin{picture}(40,20)(0,0)
\put(0,15){\line(1,0){5}}
\put(0,5){\vector(1,0){10}}
\put(15,15){\vector(-1,0){10}}
\put(15,5){\line(-1,0){5}}
\put(15,0){\framebox(10,20){}}
\put(25,15){\line(1,0){5}}
\put(25,5){\vector(1,0){10}}
\put(40,15){\vector(-1,0){10}}
\put(40,5){\line(-1,0){5}}
\end{picture}}
\put(20,10){\begin{picture}(40,20)(0,0)
\put(40,10){\oval(30,10)[r]}
\put(42,2){$\bullet$}
\put(38,-5){${\Gamma_{\hspace{-.08cm}A}}_{P_2}$}
\put(42,13){$\bullet$}
\put(38,21){${\Gamma_{\hspace{-.08cm}A}}_{P_1}$}
\put(55,10){\vector(0,1){2}}
\end{picture}}
\put(-25,10){\begin{picture}(40,20)(0,0)
\put(45,10){\oval(30,10)[l]}
\put(38,2){$\bullet$}
\put(34,-5){$\chi_{P_3}$}
\put(38,13){$\bullet$}
\put(34,21){$\chi_{P_4}$}
\put(30,10){\vector(0,-1){2}}
\end{picture}}
}
\end{picture}
-
\begin{picture}(55,30)(0,0)
\put(0,-8){
\put(20,10){\begin{picture}(40,20)(0,0)
\put(0,15){\line(1,0){5}}
\put(0,5){\vector(1,0){10}}
\put(15,15){\vector(-1,0){10}}
\put(15,5){\line(-1,0){5}}
\end{picture}}
\put(-5,10){\begin{picture}(40,20)(0,0)
\put(40,10){\oval(30,10)[r]}
\put(42,2){$\bullet$}
\put(38,-5){${\Gamma_{\hspace{-.08cm}A}}_{P_2}$}
\put(42,13){$\bullet$}
\put(38,21){${\Gamma_{\hspace{-.08cm}A}}_{P_1}$}
\put(55,10){\vector(0,1){2}}
\end{picture}}
\put(-25,10){\begin{picture}(40,20)(0,0)
\put(45,10){\oval(30,10)[l]}
\put(38,2){$\bullet$}
\put(34,-5){$\chi_{P_3}$}
\put(38,13){$\bullet$}
\put(34,21){$\chi_{P_4}$}
\put(30,10){\vector(0,-1){2}}
\end{picture}}
}
\end{picture}
\ ,
\label{ciclic 3}
\end{equation}
plus, with factor $(2 \, i \, n_\pi)^{-2}$ , (2 permutations)
%
%
\begin{equation}
\begin{picture}(55,55)(0,0)
\put(0,-8){
\put(16,13){\begin{picture}(20,40)(0,0)
\put(5,0){\line(0,1){5}}
\put(5,15){\vector(0,-1){10}}
\put(15,15){\line(0,-1){5}}
\put(15,0){\vector(0,1){10}}
\put(0,15){\framebox(20,10){}}
\put(5,25){\line(0,1){5}}
\put(5,40){\vector(0,-1){10}}
\put(15,40){\line(0,-1){5}}
\put(15,25){\vector(0,1){10}}
\end{picture}}
\put(16,8){\begin{picture}(20,40)(0,0)
\put(10,40){\oval(10,26)[t]}
\put(2,42){$\bullet$}
\put(-13,48){$\chi_{P_4}$}
\put(13,42){$\bullet$}
\put(18,48){${\Gamma_{\hspace{-.08cm}A}}_{P_1}$}
\put(10,53){\vector(-1,0){2}}
\end{picture}}
\put(16,-27){\begin{picture}(20,40)(0,0)
\put(10,45){\oval(10,26)[b]}
\put(2,38){$\bullet$}
\put(-15,34){${\Gamma_{\hspace{-.08cm}A}}_{P_3}$}
\put(13,38){$\bullet$}
\put(18,34){$\chi_{P_2}$}
\put(10,32){\vector(1,0){2}}
\end{picture}}
}
\end{picture}
+
\begin{picture}(80,30)(0,0)
\put(0,-8){
\put(20,10){\begin{picture}(40,20)(0,0)
\put(0,15){\line(1,0){5}}
\put(0,5){\vector(1,0){10}}
\put(15,15){\vector(-1,0){10}}
\put(15,5){\line(-1,0){5}}
\put(15,0){\framebox(10,20){}}
\put(25,15){\line(1,0){5}}
\put(25,5){\vector(1,0){10}}
\put(40,15){\vector(-1,0){10}}
\put(40,5){\line(-1,0){5}}
\end{picture}}
\put(20,10){\begin{picture}(40,20)(0,0)
\put(40,10){\oval(30,10)[r]}
\put(42,2){$\bullet$}
\put(38,-5){$\chi_{P_2}$}
\put(42,13){$\bullet$}
\put(38,21){${\Gamma_{\hspace{-.08cm}A}}_{P_1}$}
\put(55,10){\vector(0,1){2}}
\end{picture}}
\put(-25,10){\begin{picture}(40,20)(0,0)
\put(45,10){\oval(30,10)[l]}
\put(38,2){$\bullet$}
\put(34,-5){${\Gamma_{\hspace{-.08cm}A}}_{P_3}$}
\put(38,13){$\bullet$}
\put(34,21){$\chi_{P_4}$}
\put(30,10){\vector(0,-1){2}}
\end{picture}}
}
\end{picture}
-
\begin{picture}(55,30)(0,0)
\put(0,-8){
\put(20,10){\begin{picture}(40,20)(0,0)
\put(0,15){\line(1,0){5}}
\put(0,5){\vector(1,0){10}}
\put(15,15){\vector(-1,0){10}}
\put(15,5){\line(-1,0){5}}
\end{picture}}
\put(-5,10){\begin{picture}(40,20)(0,0)
\put(40,10){\oval(30,10)[r]}
\put(42,2){$\bullet$}
\put(38,-5){$\chi_{P_2}$}
\put(42,13){$\bullet$}
\put(38,21){${\Gamma_{\hspace{-.08cm}A}}_{P_1}$}
\put(55,10){\vector(0,1){2}}
\end{picture}}
\put(-25,10){\begin{picture}(40,20)(0,0)
\put(45,10){\oval(30,10)[l]}
\put(38,2){$\bullet$}
\put(34,-5){${\Gamma_{\hspace{-.08cm}A}}_{P_3}$}
\put(38,13){$\bullet$}
\put(34,21){$\chi_{P_4}$}
\put(30,10){\vector(0,-1){2}}
\end{picture}}
}
\end{picture}
\ .
\label{ciclic 4}
\end{equation}
I now compute these diagrams up to the order 
of $P_i \cdot P_j$ and $M_\pi^2$, starting with the
terms with move vertices $\chi_\pi$, which are closer
to the ones computed in the previous section VI.

\subsection{The 
$\chi \Gamma_{\hspace{-.08cm}A} \chi \Gamma_{\hspace{-.08cm}A}$
amplitude }

\par
Here I compute the diagrams of eq.(\ref{ciclic 4}). The first steps are identical to
the ones of Section VI, and the diagrams are identical to the ones of  
eq. (\ref{result 4 meson}) except that now the Bethe Salpeter vertex 
$\chi_{P_3}$ is replaced by the axial vertex ${\Gamma_{\hspace{-.08cm}A}}_{P_3}$,
\begin{eqnarray}
{P_4^2-M_\pi^2 \over i {\cal I}_4 } \ \
\begin{picture}(110,30)(0,0)
%
\put(15,0){
\begin{picture}(40,20)(0,0)
\put(0,15){\line(1,0){5}}
\put(0,5){\vector(1,0){10}}
\put(15,15){\vector(-1,0){10}}
\put(15,5){\line(-1,0){5}}
\put(15,0){\framebox(10,20){}}
\put(25,15){\line(1,0){5}}
\put(25,5){\vector(1,0){10}}
\put(40,15){\vector(-1,0){10}}
\put(40,5){\line(-1,0){5}}
\end{picture}}
\put(-25,0)
{\begin{picture}(40,20)(0,0)
\put(45,10){\oval(30,10)[l]}
\put(38,2){$\bullet$}
\put(34,-5){$\chi_{P_2}$}
\put(38,13){$\bullet$}
\put(34,21){${\Gamma_{\hspace{-.08cm}A}}_{P_3}$}
\put(30,10){\vector(0,-1){2}}
\end{picture}}
\put(55,0){
\begin{picture}(40,20)(0,0)
\put(0,10){\oval(10,10)[r]}
\put(3,8){$\bullet$}
\put(9,10){$ {\chi}_{P_4}{\gamma_5}_{P_1}$}
\end{picture}}
\end{picture}
+&&
\nonumber \\
{P_2^2-M_\pi^2 \over i {\cal I}_2 } \ \
\begin{picture}(110,30)(0,0)
%
\put(15,0){
\begin{picture}(40,20)(0,0)
\put(0,15){\line(1,0){5}}
\put(0,5){\vector(1,0){10}}
\put(15,15){\vector(-1,0){10}}
\put(15,5){\line(-1,0){5}}
\put(15,0){\framebox(10,20){}}
\put(25,15){\line(1,0){5}}
\put(25,5){\vector(1,0){10}}
\put(40,15){\vector(-1,0){10}}
\put(40,5){\line(-1,0){5}}
\end{picture}}
\put(-25,0)
{\begin{picture}(40,20)(0,0)
\put(45,10){\oval(30,10)[l]}
\put(38,2){$\bullet$}
\put(34,-5){${\Gamma_{\hspace{-.08cm}A}}_{P_3}$}
\put(38,13){$\bullet$}
\put(34,21){$\chi_{P_4}$}
\put(30,10){\vector(0,-1){2}}
\end{picture}}
\put(55,0){
\begin{picture}(40,20)(0,0)
\put(0,10){\oval(10,10)[r]}
\put(3,8){$\bullet$}
\put(9,10){${\gamma_5}_{P_1}{\chi}_{P_2}$}
\end{picture}}
\end{picture}
&& \ ,
\label{start chigacchiga}
\end{eqnarray}
where it is convenient to use eq. (\ref{pole pion}) to include
the ladder in the adjacent $\chi_\pi$, 
\begin{eqnarray}
&&{P_2^2-M_\pi^2 \over i {\cal I} } {P_4^2-M_\pi^2 \over i {\cal I} }
\Biggl(
\begin{picture}(150,30)(0,10)
\put(-2,0){
%
\put(45,0){
\begin{picture}(40,20)(0,0)
\put(0,15){\line(1,0){5}}
\put(0,5){\vector(1,0){10}}
\put(15,15){\vector(-1,0){10}}
\put(15,5){\line(-1,0){5}}
\put(15,0){\framebox(10,20){}}
\put(25,15){\line(1,0){5}}
\put(25,5){\vector(1,0){10}}
\put(40,15){\vector(-1,0){10}}
\put(40,5){\line(-1,0){5}}
\end{picture}}
\put(85,0){
\begin{picture}(40,20)(0,0)
\put(0,15){\line(1,0){5}}
\put(0,5){\vector(1,0){10}}
\put(15,15){\vector(-1,0){10}}
\put(15,5){\line(-1,0){5}}
\put(15,0){\framebox(10,20){}}
\put(25,15){\line(1,0){5}}
\put(25,5){\vector(1,0){10}}
\put(40,15){\vector(-1,0){10}}
\put(40,5){\line(-1,0){5}}
\end{picture}}
\put(0,0){
\begin{picture}(40,20)(0,0)
\put(45,10){\oval(10,10)[l]}
\put(37,8){$\bullet$}
\put(0,10){$\chi_{P_4} {\gamma_5}_{P_1}$}
\end{picture}}
\put(85,0){
\begin{picture}(40,20)(0,0)
\put(40,10){\oval(10,10)[r]}
\put(43,8){$\bullet$}
\put(49,10){$ \chi_{P_2} $}
\end{picture}}
\put(45,0){
\begin{picture}(40,20)(0,0)
\put(40,2){$\bullet$}
\put(40,13){$\bullet$}
\put(30,21){$ S^{-1}$}
\put(30,-6){$ {\Gamma_{\hspace{-.08cm}A}}_{P_3}$}
\end{picture}}
}
\end{picture}
\nonumber \\
&&
+
\begin{picture}(155,40)(0,0)
%
\put(25,0){
\begin{picture}(40,20)(0,0)
\put(0,15){\line(1,0){5}}
\put(0,5){\vector(1,0){10}}
\put(15,15){\vector(-1,0){10}}
\put(15,5){\line(-1,0){5}}
\put(15,0){\framebox(10,20){}}
\put(25,15){\line(1,0){5}}
\put(25,5){\vector(1,0){10}}
\put(40,15){\vector(-1,0){10}}
\put(40,5){\line(-1,0){5}}
\end{picture}}
\put(65,0){
\begin{picture}(40,20)(0,0)
\put(0,15){\line(1,0){5}}
\put(0,5){\vector(1,0){10}}
\put(15,15){\vector(-1,0){10}}
\put(15,5){\line(-1,0){5}}
\put(15,0){\framebox(10,20){}}
\put(25,15){\line(1,0){5}}
\put(25,5){\vector(1,0){10}}
\put(40,15){\vector(-1,0){10}}
\put(40,5){\line(-1,0){5}}
\end{picture}}
\put(-20,0){
\begin{picture}(40,20)(0,0)
\put(45,10){\oval(10,10)[l]}
\put(37,8){$\bullet$}
\put(20,10){$\chi_{P_4}$}
\end{picture}}
\put(65,0){
\begin{picture}(40,20)(0,0)
\put(40,10){\oval(10,10)[r]}
\put(43,8){$\bullet$}
\put(49,10){$  {\gamma_5}_{P_1} \chi_{P_2}$}
\end{picture}}
\put(25,0){
\begin{picture}(40,20)(0,0)
\put(40,2){$\bullet$}
\put(40,13){$\bullet$}
\put(30,21){$ S^{-1}$}
\put(30,-6){$ {\Gamma_{\hspace{-.08cm}A}}_{P_3}$}
\end{picture}}
\end{picture}
\Biggr)
\ ,
\label{inter chigachiga}
\end{eqnarray}
and to apply the Ward identity (\ref{crucial}) to the axial 
vertex ${\Gamma_{\hspace{-.08cm}A}}_{P_3}$, 
\begin{eqnarray}
&{P_4^2-M_\pi^2 \over i {\cal I} }
\begin{picture}(140,20)(0,0)
%
\put(65,0){
\begin{picture}(40,20)(0,0)
\put(0,15){\line(1,0){5}}
\put(0,5){\vector(1,0){10}}
\put(15,15){\vector(-1,0){10}}
\put(15,5){\line(-1,0){5}}
\end{picture}}
\put(20,0){
\begin{picture}(40,20)(0,0)
\put(45,10){\oval(10,10)[l]}
\put(37,8){$\bullet$}
\put(-20,10){${\gamma_5}_{P_3}\chi_{P_4}{\gamma_5}_{P_1} $}
\end{picture}}
\put(40,0){
\begin{picture}(40,20)(0,0)
\put(40,10){\oval(10,10)[r]}
\put(43,8){$\bullet$}
\put(49,10){${\chi}_{P_2} $}
\end{picture}}
\end{picture}
&
\nonumber \\
&+
{P_2^2-M_\pi^2 \over i {\cal I} } {P_4^2-M_\pi^2 \over i {\cal I} }
\begin{picture}(140,20)(0,0)
%
\put(45,0){
\begin{picture}(40,20)(0,0)
\put(0,15){\line(1,0){5}}
\put(0,5){\vector(1,0){10}}
\put(15,15){\vector(-1,0){10}}
\put(15,5){\line(-1,0){5}}
\put(15,0){\framebox(10,20){}}
\put(25,15){\line(1,0){5}}
\put(25,5){\vector(1,0){10}}
\put(40,15){\vector(-1,0){10}}
\put(40,5){\line(-1,0){5}}
\end{picture}}
\put(0,0){
\begin{picture}(40,20)(0,0)
\put(45,10){\oval(10,10)[l]}
\put(37,8){$\bullet$}
\put(0,10){$\chi_{P_4}{\gamma_5}_{P_1} $}
\end{picture}}
\put(45,0){
\begin{picture}(40,20)(0,0)
\put(40,10){\oval(10,10)[r]}
\put(43,8){$\bullet$}
\put(49,10){${\chi}_{P_2}{\gamma_5}_{P_3}$}
\end{picture}}
\end{picture}
&
\nonumber \\
&+
{P_2^2-M_\pi^2 \over i {\cal I} } 
\begin{picture}(140,20)(0,0)
%
\put(25,0){
\begin{picture}(40,20)(0,0)
\put(0,15){\line(1,0){5}}
\put(0,5){\vector(1,0){10}}
\put(15,15){\vector(-1,0){10}}
\put(15,5){\line(-1,0){5}}
\end{picture}}
\put(-20,0){
\begin{picture}(40,20)(0,0)
\put(45,10){\oval(10,10)[l]}
\put(37,8){$\bullet$}
\put(20,10){$\chi_{P_4} $}
\end{picture}}
\put(0,0){
\begin{picture}(40,20)(0,0)
\put(40,10){\oval(10,10)[r]}
\put(43,8){$\bullet$}
\put(49,10){$ {\gamma_5}_{P_1}\chi_{P_2}{\gamma_5}_{P_3}$}
\end{picture}}
\end{picture}
&
\nonumber \\
&+
{P_2^2-M_\pi^2 \over i {\cal I} } {P_4^2-M_\pi^2 \over i {\cal I} }
\begin{picture}(140,20)(0,0)
%
\put(45,0){
\begin{picture}(40,20)(0,0)
\put(0,15){\line(1,0){5}}
\put(0,5){\vector(1,0){10}}
\put(15,15){\vector(-1,0){10}}
\put(15,5){\line(-1,0){5}}
\put(15,0){\framebox(10,20){}}
\put(25,15){\line(1,0){5}}
\put(25,5){\vector(1,0){10}}
\put(40,15){\vector(-1,0){10}}
\put(40,5){\line(-1,0){5}}
\end{picture}}
\put(0,0){
\begin{picture}(40,20)(0,0)
\put(45,10){\oval(10,10)[l]}
\put(37,8){$\bullet$}
\put(0,10){${\gamma_5}_{P_3}\chi_{P_4} $}
\end{picture}}
\put(45,0){
\begin{picture}(40,20)(0,0)
\put(40,10){\oval(10,10)[r]}
\put(43,8){$\bullet$}
\put(49,10){$ {\gamma_5}_{P_1}\chi_{P_2}$}
\end{picture}}
\end{picture}
& \ ,
\label{least chigachiga}
\end{eqnarray}
where this result is exact. Only the 
contribution up to order second order $P_i \cdot P_j$ or
$M_\pi^2$ needs to be retained. In eq. (\ref{least chigachiga}) there
are two classes of diagrams. The second and fourth diagrams 
have scalar or vector vertices $\chi\gamma_5$, and therefore 
the ladder is not a pseudoscalar and is not able able to
cancel the higher order factor 
${P_2^2-M_\pi^2 \over i {\cal I} } {P_4^2-M_\pi^2 \over i {\cal I} }$.
The first and third diagrams, with a pseudoscalar vertex $\chi$,
already have a factor ${P^2-M_\pi^2 \over i {\cal I} }$ of second
order. Therefore the remaining trace $tr\{\gamma_5  \chi \gamma_5 
 S \chi S \}$ only needs to be computed at order zero, where this trace
is simply the constant  ${\cal I}$ which was defined in eq. (\ref{off mass shell}).
The final result up to second order is,

\begin{equation}
{ P_2^2 -M_\pi^2 + P_4^2 -M_\pi^2 \over i}
\label{last ciclic 4}
\end{equation}
%

\subsection{The 
$\chi \chi \Gamma_{\hspace{-.08cm}A} \Gamma_{\hspace{-.08cm}A}$
amplitude }

\par
Here I compute the diagrams of eq.(\ref{ciclic 3}). Again the first steps are similar to
the ones of Section VI, and the diagrams are identical to the ones of  
eq. (\ref{result 4 meson}) except that the Bethe Salpeter vertex 
$\chi_{P_2}$ is replaced by the axial vertex ${\Gamma_{\hspace{-.08cm}A}}_{P_2}$.
In the second diagram of eq. eq.(\ref{ciclic 3}), the axial vertex  
${\Gamma_{\hspace{-.08cm}A}}_{P_2}$ is adjacent to the vertex 
${\Gamma_{\hspace{-.08cm}A}}_{P_1}$ and eq. (\ref{second}) substitutes 
eq. (\ref{pole pion}). The equation similar to eq. (\ref{result 4 meson})
is now,
\begin{eqnarray}
&
{P_4^2-M_\pi^2 \over i {\cal I}}
\
\begin{picture}(115,30)(0,0)
%
\put(20,0){
\begin{picture}(40,20)(0,0)
\put(0,15){\line(1,0){5}}
\put(0,5){\vector(1,0){10}}
\put(15,15){\vector(-1,0){10}}
\put(15,5){\line(-1,0){5}}
\put(15,0){\framebox(10,20){}}
\put(25,15){\line(1,0){5}}
\put(25,5){\vector(1,0){10}}
\put(40,15){\vector(-1,0){10}}
\put(40,5){\line(-1,0){5}}
\end{picture}}
\put(-25,0){
\begin{picture}(40,20)(0,0)
\put(45,10){\oval(30,10)[l]}
\put(38,2){$\bullet$}
\put(34,21){$\chi_{P_3}$}
\put(38,13){$\bullet$}
\put(34,-6){${\Gamma_{\hspace{-.08cm}A}}_{P_2}$}
\put(30,10){\vector(0,-1){2}}
\end{picture}}
\put(20,0){
\begin{picture}(40,20)(0,0)
\put(40,10){\oval(10,10)[r]}
\put(43,8){$\bullet$}
\put(49,10){$ \chi_{P_4} {\gamma_5}_{P_1} $}
\end{picture}}
\end{picture}
&
\nonumber \\
&
+
\begin{picture}(115,30)(0,0)
%
\put(20,0){
\begin{picture}(40,20)(0,0)
\put(0,15){\line(1,0){5}}
\put(0,5){\vector(1,0){10}}
\put(15,15){\vector(-1,0){10}}
\put(15,5){\line(-1,0){5}}
\put(15,0){\framebox(10,20){}}
\put(25,15){\line(1,0){5}}
\put(25,5){\vector(1,0){10}}
\put(40,15){\vector(-1,0){10}}
\put(40,5){\line(-1,0){5}}
\end{picture}}
\put(-25,0){
\begin{picture}(40,20)(0,0)
\put(45,10){\oval(30,10)[l]}
\put(38,2){$\bullet$}
\put(34,21){$\chi_{P_4}$}
\put(38,13){$\bullet$}
\put(34,-3){$\chi_{P_3}$}
\put(30,10){\vector(0,-1){2}}
\end{picture}}
\put(20,0){
\begin{picture}(40,20)(0,0)
\put(40,10){\oval(10,10)[r]}
\put(43,8){$\bullet$}
\put(49,10){$ {\gamma_5}_{P_1} {\gamma_{\hspace{-.08cm}A}}_{P_2} $}
\end{picture}}
\end{picture}
& \,
\label{first ccgg}
\end{eqnarray}
where the last term is already proportional to 
$ {\gamma_{\hspace{-.08cm}A}}_{P_2} $
so it only needs one of the $\chi$ to produce
a term of second order . So in the last term it is 
convenient to return to a lower order in
$\chi -{i \over 2 f_\pi} \Gamma_{\hspace{-.08cm}A} $,
where $\chi \chi$ is be replaced by
$\chi {i \over 2 f_\pi} \Gamma_{\hspace{-.08cm}A} 
+{i \over 2 f_\pi} \Gamma_{\hspace{-.08cm}A}\chi 
- {i \over 2 f_\pi} \Gamma_{\hspace{-.08cm}A}
{i \over 2 f_\pi} \Gamma_{\hspace{-.08cm}A} $. 
Including the desired
ladders, there are four different terms,
\FL
\begin{eqnarray}
&
=
&
\begin{picture}(70,10)(0,0)
\put(5,2){${ {P_3^2-M_\pi^2 \over i {\cal I}} 
{P_4^2-M_\pi^2 \over i {\cal I}}}$}
\end{picture}
%
\begin{picture}(155,30)(0,0)
%
\put(25,0){
\begin{picture}(40,20)(0,0)
\put(0,15){\line(1,0){5}}
\put(0,5){\vector(1,0){10}}
\put(15,15){\vector(-1,0){10}}
\put(15,5){\line(-1,0){5}}
\put(15,0){\framebox(10,20){}}
\put(25,15){\line(1,0){5}}
\put(25,5){\vector(1,0){10}}
\put(40,15){\vector(-1,0){10}}
\put(40,5){\line(-1,0){5}}
\end{picture}}
\put(65,0){
\begin{picture}(40,20)(0,0)
\put(0,15){\line(1,0){5}}
\put(0,5){\vector(1,0){10}}
\put(15,15){\vector(-1,0){10}}
\put(15,5){\line(-1,0){5}}
\put(15,0){\framebox(10,20){}}
\put(25,15){\line(1,0){5}}
\put(25,5){\vector(1,0){10}}
\put(40,15){\vector(-1,0){10}}
\put(40,5){\line(-1,0){5}}
\end{picture}}
\put(-20,0){
\begin{picture}(40,20)(0,0)
\put(45,10){\oval(10,10)[l]}
\put(37,8){$\bullet$}
\put(20,10){$\chi_{P_3}$}
\end{picture}}
\put(65,0){
\begin{picture}(40,20)(0,0)
\put(40,10){\oval(10,10)[r]}
\put(43,8){$\bullet$}
\put(49,10){$ \chi_{P_4} {\gamma_5}_{P_1}$}
\end{picture}}
\put(25,0){
\begin{picture}(40,20)(0,0)
\put(40,2){$\bullet$}
\put(40,13){$\bullet$}
\put(30,21){$ S^{-1}$}
\put(30,-6){$ {\Gamma_{\hspace{-.08cm}A}}_{P_2}$}
\end{picture}}
\end{picture}
\nonumber \\
&
&
+
{i \over 2 f_\pi}
{P_4^2-M_\pi^2 \over  i {\cal I}}
\begin{picture}(155,40)(0,0)
%
\put(25,0){
\begin{picture}(40,20)(0,0)
\put(0,15){\line(1,0){5}}
\put(0,5){\vector(1,0){10}}
\put(15,15){\vector(-1,0){10}}
\put(15,5){\line(-1,0){5}}
\put(15,0){\framebox(10,20){}}
\put(25,15){\line(1,0){5}}
\put(25,5){\vector(1,0){10}}
\put(40,15){\vector(-1,0){10}}
\put(40,5){\line(-1,0){5}}
\end{picture}}
\put(65,0){
\begin{picture}(40,20)(0,0)
\put(0,15){\line(1,0){5}}
\put(0,5){\vector(1,0){10}}
\put(15,15){\vector(-1,0){10}}
\put(15,5){\line(-1,0){5}}
\put(15,0){\framebox(10,20){}}
\put(25,15){\line(1,0){5}}
\put(25,5){\vector(1,0){10}}
\put(40,15){\vector(-1,0){10}}
\put(40,5){\line(-1,0){5}}
\end{picture}}
\put(-20,0){
\begin{picture}(40,20)(0,0)
\put(45,10){\oval(10,10)[l]}
\put(37,8){$\bullet$}
\put(20,10){$\chi_{P_4}$}
\end{picture}}
\put(65,0){
\begin{picture}(40,20)(0,0)
\put(40,10){\oval(10,10)[r]}
\put(43,8){$\bullet$}
\put(49,10){${\gamma_5}_{P_1} {\gamma_{\hspace{-.08cm}A}}_{P_2}$}
\end{picture}}
\put(25,0){
\begin{picture}(40,20)(0,0)
\put(40,2){$\bullet$}
\put(40,13){$\bullet$}
\put(30,21){$ S^{-1}$}
\put(30,-6){$ {\Gamma_{\hspace{-.08cm}A}}_{P_3}$}
\end{picture}}
\end{picture}
\nonumber \\
&
&
+
{P_3^2-M_\pi^2 \over i {\cal I}}
{i \over 2 f_\pi}
\begin{picture}(155,40)(0,0)
%
\put(25,0){
\begin{picture}(40,20)(0,0)
\put(0,15){\line(1,0){5}}
\put(0,5){\vector(1,0){10}}
\put(15,15){\vector(-1,0){10}}
\put(15,5){\line(-1,0){5}}
\put(15,0){\framebox(10,20){}}
\put(25,15){\line(1,0){5}}
\put(25,5){\vector(1,0){10}}
\put(40,15){\vector(-1,0){10}}
\put(40,5){\line(-1,0){5}}
\end{picture}}
\put(65,0){
\begin{picture}(40,20)(0,0)
\put(0,15){\line(1,0){5}}
\put(0,5){\vector(1,0){10}}
\put(15,15){\vector(-1,0){10}}
\put(15,5){\line(-1,0){5}}
\put(15,0){\framebox(10,20){}}
\put(25,15){\line(1,0){5}}
\put(25,5){\vector(1,0){10}}
\put(40,15){\vector(-1,0){10}}
\put(40,5){\line(-1,0){5}}
\end{picture}}
\put(-20,0){
\begin{picture}(40,20)(0,0)
\put(45,10){\oval(10,10)[l]}
\put(37,8){$\bullet$}
\put(20,10){$\chi_{P_3}$}
\end{picture}}
\put(65,0){
\begin{picture}(40,20)(0,0)
\put(40,10){\oval(10,10)[r]}
\put(43,8){$\bullet$}
\put(49,10){$ {\gamma_5}_{P_1} {\gamma_{\hspace{-.08cm}A}}_{P_2}$}
\end{picture}}
\put(25,0){
\begin{picture}(40,20)(0,0)
\put(40,2){$\bullet$}
\put(40,13){$\bullet$}
\put(30,-6){$ S^{-1}$}
\put(30,21){$ {\Gamma_{\hspace{-.08cm}A}}_{P_4}$}
\end{picture}}
\end{picture}
\nonumber \\
&
&
-
{i \over 2 f_\pi}
{i \over 2 f_\pi}
\begin{picture}(155,40)(0,0)
%
\put(25,0){
\begin{picture}(40,20)(0,0)
\put(0,15){\line(1,0){5}}
\put(0,5){\vector(1,0){10}}
\put(15,15){\vector(-1,0){10}}
\put(15,5){\line(-1,0){5}}
\put(15,0){\framebox(10,20){}}
\put(25,15){\line(1,0){5}}
\put(25,5){\vector(1,0){10}}
\put(40,15){\vector(-1,0){10}}
\put(40,5){\line(-1,0){5}}
\end{picture}}
\put(65,0){
\begin{picture}(40,20)(0,0)
\put(0,15){\line(1,0){5}}
\put(0,5){\vector(1,0){10}}
\put(15,15){\vector(-1,0){10}}
\put(15,5){\line(-1,0){5}}
\put(15,0){\framebox(10,20){}}
\put(25,15){\line(1,0){5}}
\put(25,5){\vector(1,0){10}}
\put(40,15){\vector(-1,0){10}}
\put(40,5){\line(-1,0){5}}
\end{picture}}
\put(-20,0){
\begin{picture}(40,20)(0,0)
\put(45,10){\oval(10,10)[l]}
\put(37,8){$\bullet$}
\put(20,10){${\gamma_{\hspace{-.08cm}A}}_{P_4}$}
\end{picture}}
\put(65,0){
\begin{picture}(40,20)(0,0)
\put(40,10){\oval(10,10)[r]}
\put(43,8){$\bullet$}
\put(49,10){$ {\gamma_5}_{P_1} {\gamma_{\hspace{-.08cm}A}}_{P_2}$}
\end{picture}}
\put(25,0){
\begin{picture}(40,20)(0,0)
\put(40,2){$\bullet$}
\put(40,13){$\bullet$}
\put(30,21){$ S^{-1}$}
\put(30,-6){$ {\Gamma_{\hspace{-.08cm}A}}_{P_3}$}
\end{picture}}
\end{picture}
.
\end{eqnarray}
Using again the Ward identity (\ref{crucial}), and removing the
higher order cases where the ladder is scalar, eq. (\ref{first ccgg}) 
simplifies to,
\begin{eqnarray}
=
&
{P_3^2-M_\pi^2 \over i {\cal I}} 
{P_4^2-M_\pi^2 \over i {\cal I}}
\
\begin{picture}(135,30)(0,0)
%
\put(25,0){
\begin{picture}(40,20)(0,0)
\put(0,15){\line(1,0){5}}
\put(0,5){\vector(1,0){10}}
\put(15,15){\vector(-1,0){10}}
\put(15,5){\line(-1,0){5}}
\put(15,0){\framebox(10,20){}}
\put(25,15){\line(1,0){5}}
\put(25,5){\vector(1,0){10}}
\put(40,15){\vector(-1,0){10}}
\put(40,5){\line(-1,0){5}}
\end{picture}}
\put(-20,0){
\begin{picture}(40,20)(0,0)
\put(45,10){\oval(10,10)[l]}
\put(37,8){$\bullet$}
\put(20,10){$\chi_{P_3}$}
\end{picture}}
\put(25,0){
\begin{picture}(40,20)(0,0)
\put(40,10){\oval(10,10)[r]}
\put(43,8){$\bullet$}
\put(49,10){$ \chi_{P_4} {\gamma_5}_{P_1} {\gamma_5}_{P_2}$}
\end{picture}}
\end{picture}
&
\nonumber \\
&
+
{i \over 2 f_\pi}
{P_4^2-M_\pi^2 \over  i {\cal I}}
\
\begin{picture}(135,30)(0,0)
%
\put(25,0){
\begin{picture}(40,20)(0,0)
\put(0,15){\line(1,0){5}}
\put(0,5){\vector(1,0){10}}
\put(15,15){\vector(-1,0){10}}
\put(15,5){\line(-1,0){5}}
\put(15,0){\framebox(10,20){}}
\put(25,15){\line(1,0){5}}
\put(25,5){\vector(1,0){10}}
\put(40,15){\vector(-1,0){10}}
\put(40,5){\line(-1,0){5}}
\end{picture}}
\put(-20,0){
\begin{picture}(40,20)(0,0)
\put(45,10){\oval(10,10)[l]}
\put(37,8){$\bullet$}
\put(20,10){$\chi_{P_4}$}
\end{picture}}
\put(25,0){
\begin{picture}(40,20)(0,0)
\put(40,10){\oval(10,10)[r]}
\put(43,8){$\bullet$}
\put(49,10){${\gamma_5}_{P_1} {\gamma_{\hspace{-.08cm}A}}_{P_2}{\gamma_5}_{P_3}$}
\end{picture}}
\end{picture}
&
\nonumber \\
&
+
{P_3^2-M_\pi^2 \over i {\cal I}}
{i \over 2 f_\pi}
\
\begin{picture}(135,30)(0,0)
%
\put(25,0){
\begin{picture}(40,20)(0,0)
\put(0,15){\line(1,0){5}}
\put(0,5){\vector(1,0){10}}
\put(15,15){\vector(-1,0){10}}
\put(15,5){\line(-1,0){5}}
\put(15,0){\framebox(10,20){}}
\put(25,15){\line(1,0){5}}
\put(25,5){\vector(1,0){10}}
\put(40,15){\vector(-1,0){10}}
\put(40,5){\line(-1,0){5}}
\end{picture}}
\put(-20,0){
\begin{picture}(40,20)(0,0)
\put(45,10){\oval(10,10)[l]}
\put(37,8){$\bullet$}
\put(20,10){$\chi_{P_3}$}
\end{picture}}
\put(25,0){
\begin{picture}(40,20)(0,0)
\put(40,10){\oval(10,10)[r]}
\put(43,8){$\bullet$}
\put(49,10){$ {\gamma_5}_{P_4} {\gamma_5}_{P_1} {\gamma_{\hspace{-.08cm}A}}_{P_2}$}
\end{picture}}
\end{picture}
&
\nonumber \\
&
-
{i \over 2 f_\pi}
{i \over 2 f_\pi}
\
\begin{picture}(135,30)(0,0)
%
\put(25,0){
\begin{picture}(40,20)(0,0)
\put(0,15){\line(1,0){5}}
\put(0,5){\vector(1,0){10}}
\put(15,15){\vector(-1,0){10}}
\put(15,5){\line(-1,0){5}}
\put(15,0){\framebox(10,20){}}
\put(25,15){\line(1,0){5}}
\put(25,5){\vector(1,0){10}}
\put(40,15){\vector(-1,0){10}}
\put(40,5){\line(-1,0){5}}
\end{picture}}
\put(-20,0){
\begin{picture}(40,20)(0,0)
\put(45,10){\oval(10,10)[l]}
\put(37,8){$\bullet$}
\put(20,10){${\gamma_{\hspace{-.08cm}A}}_{P_4}$}
\end{picture}}
\put(25,0){
\begin{picture}(40,20)(0,0)
\put(40,10){\oval(10,10)[r]}
\put(43,8){$\bullet$}
\put(49,10){$ {\gamma_5}_{P_1} {\gamma_{\hspace{-.08cm}A}}_{P_2} {\gamma_5}_{P_3}$}
\end{picture}}
\end{picture}
\ ,
&
\end{eqnarray}
where the ladders can all be re absorbed in vertices,
\begin{eqnarray}
=
&
{P_4^2-M_\pi^2 \over i {\cal I}}
tr\{(S \chi S)_{P_3} \chi_{P_4} \}
&
\nonumber \\
&
+
{i \over 2 f_\pi}
tr\{(S \chi S)_{P_4}  
{\gamma_{\hspace{-.08cm}A}}_{-P_2}\}
&
\nonumber \\
&
+
{i \over 2 f_\pi}
tr\{(S \chi S)_{P_3}  
{\gamma_{\hspace{-.08cm}A}}_{P_2}\}
&
\nonumber \\
&
-
{i \over 2 f_\pi}
{i \over 2 f_\pi}
tr\{(S{\Gamma_{\hspace{-.08cm}A}}S)_{P_4} {\gamma_{\hspace{-.08cm}A}}_{-P_2}\}
&
\nonumber \\
=
&
{P_4^2-M_\pi^2 \over i } + {P_4 \cdot P_2-M_\pi^2 \over i } 
+ {-P_3 \cdot P_2 -M_\pi^2 \over i } -{ -M_\pi^2 \over i }
&
\nonumber \\
=
&
{P_3^2 + P_4^2 +P_3 \cdot P_4 
+P_3 \cdot P_1 +P_4 \cdot P_2 -2 \, M_\pi^2 \over i }
\ ,
&
\label{last ciclic 3}
\end{eqnarray}
and this is the final result of eq. (\ref{ciclic 3}) up to second order.

%
\par
\subsection{ The 
$\chi \Gamma_{\hspace{-.08cm}A} \Gamma_{\hspace{-.08cm}A} \Gamma_{\hspace{-.08cm}A}$
amplitude }

\par
I now compute the diagrams of eq.(\ref{ciclic 2}). The first step is
identical to the previous case except that the Bethe Salpeter vertex 
$\chi_{P_3}$ is replaced by the axial vertex ${\Gamma_{\hspace{-.08cm}A}}_{P_3}$.
The equation similar to eq. (\ref{result 4 meson}) is now,
\begin{eqnarray}
&
{P_4^2-M_\pi^2 \over i {\cal I}}
\
\begin{picture}(115,30)(0,0)
%
\put(20,0){
\begin{picture}(40,20)(0,0)
\put(0,15){\line(1,0){5}}
\put(0,5){\vector(1,0){10}}
\put(15,15){\vector(-1,0){10}}
\put(15,5){\line(-1,0){5}}
\put(15,0){\framebox(10,20){}}
\put(25,15){\line(1,0){5}}
\put(25,5){\vector(1,0){10}}
\put(40,15){\vector(-1,0){10}}
\put(40,5){\line(-1,0){5}}
\end{picture}}
\put(-25,0){
\begin{picture}(40,20)(0,0)
\put(45,10){\oval(30,10)[l]}
\put(38,2){$\bullet$}
\put(34,21){${\Gamma_{\hspace{-.08cm}A}}_{P_3}$}
\put(38,13){$\bullet$}
\put(34,-6){${\Gamma_{\hspace{-.08cm}A}}_{P_2}$}
\put(30,10){\vector(0,-1){2}}
\end{picture}}
\put(20,0){
\begin{picture}(40,20)(0,0)
\put(40,10){\oval(10,10)[r]}
\put(43,8){$\bullet$}
\put(49,10){$ \chi_{P_4} {\gamma_5}_{P_1} $}
\end{picture}}
\end{picture}
&
\nonumber \\
&
+
\begin{picture}(115,30)(0,0)
%
\put(20,0){
\begin{picture}(40,20)(0,0)
\put(0,15){\line(1,0){5}}
\put(0,5){\vector(1,0){10}}
\put(15,15){\vector(-1,0){10}}
\put(15,5){\line(-1,0){5}}
\put(15,0){\framebox(10,20){}}
\put(25,15){\line(1,0){5}}
\put(25,5){\vector(1,0){10}}
\put(40,15){\vector(-1,0){10}}
\put(40,5){\line(-1,0){5}}
\end{picture}}
\put(-25,0){
\begin{picture}(40,20)(0,0)
\put(45,10){\oval(30,10)[l]}
\put(38,2){$\bullet$}
\put(34,21){$\chi_{P_4}$}
\put(38,13){$\bullet$}
\put(34,-6){${\Gamma_{\hspace{-.08cm}A}}_{P_3}$}
\put(30,10){\vector(0,-1){2}}
\end{picture}}
\put(20,0){
\begin{picture}(40,20)(0,0)
\put(40,10){\oval(10,10)[r]}
\put(43,8){$\bullet$}
\put(49,10){$ {\gamma_5}_{P_1} {\gamma_{\hspace{-.08cm}A}}_{P_2} $}
\end{picture}}
\end{picture}
& \ ,
\label{first cggg}
\end{eqnarray}
and the Ward identity (\ref{crucial}) can now be used in the axial Vertex 
${\Gamma_{\hspace{-.08cm}A}}_{P_3}$. Again eq.
(\ref{pole pion}) or eq. (\ref{second}) are needed to introduce a second ladder
in the loop. Excluding the terms with vertex scalar or vector vertex
$\chi \gamma_5$ which have a higher order, eq. (\ref{first cggg}) simplifies to,
\begin{eqnarray}
&
{P_4^2-M_\pi^2 \over i {\cal I}}
\
\begin{picture}(120,15)(0,0)
\put(25,-5){
\begin{picture}(40,20)(0,0)
\put(0,15){\line(1,0){5}}
\put(0,5){\vector(1,0){10}}
\put(15,15){\vector(-1,0){10}}
\put(15,5){\line(-1,0){5}}
\end{picture}}
\put(-5,-5){
\begin{picture}(40,20)(0,0)
\put(30,10){\oval(10,10)[l]}
\put(22,8){$\bullet$}
\put(2,10){${\Gamma_{\hspace{-.08cm}A}}_{P_2}$}
\end{picture}}
\put(0,-5){
\begin{picture}(40,20)(0,0)
\put(40,10){\oval(10,10)[r]}
\put(43,8){$\bullet$}
\put(50,10){$
{\gamma_5 }_{P_3}\chi_{P_4}{\gamma_5 }_{P_1}
$}
\end{picture}}
\end{picture}
&
\nonumber \\
&
+ 
\begin{picture}(120,15)(0,0)
\put(25,-5){
\begin{picture}(40,20)(0,0)
\put(0,15){\line(1,0){5}}
\put(0,5){\vector(1,0){10}}
\put(15,15){\vector(-1,0){10}}
\put(15,5){\line(-1,0){5}}
\end{picture}}
\put(-5,-5){
\begin{picture}(40,20)(0,0)
\put(30,10){\oval(10,10)[l]}
\put(22,8){$\bullet$}
\put(5,10){$\chi_{P_4}$}
\end{picture}}
\put(0,-5){
\begin{picture}(40,20)(0,0)
\put(40,10){\oval(10,10)[r]}
\put(43,8){$\bullet$}
\put(50,10){$
{\gamma_5 }_{P_1}{\gamma_{\hspace{-.08cm}A}}_{P_2}{\gamma_5 }_{P_3}
$}
\end{picture}}
\end{picture}
& \ ,
\label{quasi chigagaga}
\end{eqnarray}
and this can be further simplified when only the second order is retained.
Using eq. (\ref{chi = ga}) the order zero of the loop in the first diagram  
simplifies to $2 \, i \, n_\pi \cal I$. The second loop is computed with
the trace ({condensed}). The result for eq. (\ref{ciclic 2}) is,
\begin{equation}
(2 \, n_\pi) \left[ \left( P_4+P_2 \right) \cdot P_4 -2 M_\pi^2 \right] \ .
\label{last ciclic 2}
\end{equation}

%
%
\par
\subsection{ Four $\Gamma_A$ }

\par
I now compute the diagrams of eq. (\ref{ciclic 1}), where all the vertices
are axial vertices ${\Gamma_{\hspace{-.08cm}A}}_{P_i}$. The technique to simplify
the loops is identical to the previous cases, but  
the term with a scalar or vector ladder must also be considered
because it contributes to the second order.
The number of vertices is again decreased with the 
help of eqs. (\ref{crucial}) and (\ref{second}),
%
%
\begin{eqnarray}
&
\begin{picture}(80,30)(0,0)
\put(0,-8){
\put(5,10){
\begin{picture}(40,20)(0,0)
\put(0,15){\line(1,0){5}}
\put(0,5){\vector(1,0){10}}
\put(15,15){\vector(-1,0){10}}
\put(15,5){\line(-1,0){5}}
\put(15,0){\framebox(10,20){}}
\put(25,15){\line(1,0){5}}
\put(25,5){\vector(1,0){10}}
\put(40,15){\vector(-1,0){10}}
\put(40,5){\line(-1,0){5}}
\put(40,10){\oval(30,10)[r]}
\put(42,2){$\bullet$}
\put(48,-6){${\Gamma_{\hspace{-.08cm}A}}_{P_2}$}
\put(42,13){$\bullet$}
\put(48,18){${\Gamma_{\hspace{-.08cm}A}}_{P_1}$}
\put(55,10){\vector(0,1){2}}
\end{picture}
}
}
\end{picture}
=&
\begin{picture}(115,30)(0,0)
\put(0,-8){
\put(5,10){\begin{picture}(40,20)(0,0)
\put(0,15){\line(1,0){5}}
\put(0,5){\vector(1,0){10}}
\put(15,15){\vector(-1,0){10}}
\put(15,5){\line(-1,0){5}}
\put(15,0){\framebox(10,20){}}
\put(25,15){\line(1,0){5}}
\put(25,5){\vector(1,0){10}}
\put(40,15){\vector(-1,0){10}}
\put(40,5){\line(-1,0){5}}
\end{picture}}
\put(5,10){\begin{picture}(60,20)(0,0)
\put(38,2){$\bullet$}
\put(35,-8){$S^{-1}$}
\put(38,13){$\bullet$}
\put(33,22){${\Gamma_{\hspace{-.08cm}A}}_{P_1}$}
\end{picture}}
%
\put(45,10){\begin{picture}(40,20)(0,0)
\put(0,15){\line(1,0){5}}
\put(0,5){\vector(1,0){10}}
\put(15,15){\vector(-1,0){10}}
\put(15,5){\line(-1,0){5}}
\put(15,0){\framebox(10,20){}}
\put(25,15){\line(1,0){5}}
\put(25,5){\vector(1,0){10}}
\put(40,15){\vector(-1,0){10}}
\put(40,5){\line(-1,0){5}}
\end{picture}}
\put(45,10){\begin{picture}(40,20)(0,0)
\put(40,10){\oval(10,10)[r]}
\put(43,8){$\bullet$}
\put(49,10){${\gamma_{\hspace{-.08cm}A}}_{P_2}$}
\end{picture}}
}
\end{picture}
\label{step}
\nonumber \\
&=
\begin{picture}(65,20)(0,0)
\put(0,-13){
\put(-5,5){\begin{picture}(40,20)(0,0)
\put(25,15){\line(1,0){5}}
\put(25,5){\vector(1,0){10}}
\put(40,15){\vector(-1,0){10}}
\put(40,5){\line(-1,0){5}}
\end{picture}}
\put(-5,5){\begin{picture}(40,20)(0,0)
\put(40,10){\oval(10,10)[r]}
\put(43,8){$\bullet$}
\put(6,15){${\gamma_5}_{P_1}$}
\put(49,10){${\Gamma_{\hspace{-.08cm}A}}_{P_2}$}
\end{picture}}
}
\end{picture}
+&
\begin{picture}(95,20)(0,0)
\put(0,-13){
\put(5,5){\begin{picture}(40,20)(0,0)
\put(0,15){\line(1,0){5}}
\put(0,5){\vector(1,0){10}}
\put(15,15){\vector(-1,0){10}}
\put(15,5){\line(-1,0){5}}
\put(15,0){\framebox(10,20){}}
\put(25,15){\line(1,0){5}}
\put(25,5){\vector(1,0){10}}
\put(40,15){\vector(-1,0){10}}
\put(40,5){\line(-1,0){5}}
\end{picture}}
\put(5,5){\begin{picture}(40,20)(0,0)
\put(40,10){\oval(10,10)[r]}
\put(43,8){$\bullet$}
\put(49,10){${\gamma_5}_{P_1}{\gamma_{\hspace{-.08cm}A}}_{P_2}$}
\end{picture}}
}
\end{picture}
\ ,
\end{eqnarray}
where the Ward Identity (\ref{crucial}) can also be applied
to the vertex ${\Gamma_{\hspace{-.08cm}A}}_{P_2}$, rather 
than be applied to the vertex ${\Gamma_{\hspace{-.08cm}A}}_{P_1}$. 
It is more convenient to take the average of these two possible choices. 
This is repeated with the vertices ${\Gamma_{\hspace{-.08cm}A}}_{P_3}$ 
and ${\Gamma_{\hspace{-.08cm}A}}_{P_4}$ to compute the square box with ladder,
%
%
\begin{eqnarray}
&&
\begin{picture}(80,25)(0,0)
\put(0,-15){
\put(20,10){\begin{picture}(40,20)(0,0)
\put(0,15){\line(1,0){5}}
\put(0,5){\vector(1,0){10}}
\put(15,15){\vector(-1,0){10}}
\put(15,5){\line(-1,0){5}}
\put(15,0){\framebox(10,20){}}
\put(25,15){\line(1,0){5}}
\put(25,5){\vector(1,0){10}}
\put(40,15){\vector(-1,0){10}}
\put(40,5){\line(-1,0){5}}
\end{picture}}
\put(20,10){\begin{picture}(40,20)(0,0)
\put(40,10){\oval(30,10)[r]}
\put(42,2){$\bullet$}
\put(38,-5){${\Gamma_{\hspace{-.08cm}A}}_{P_2}$}
\put(42,13){$\bullet$}
\put(38,21){${\Gamma_{\hspace{-.08cm}A}}_{P_1}$}
\put(55,10){\vector(0,1){2}}
\end{picture}}
\put(-25,10){\begin{picture}(40,20)(0,0)
\put(45,10){\oval(30,10)[l]}
\put(38,2){$\bullet$}
\put(34,-5){${\Gamma_{\hspace{-.08cm}A}}_{P_3}$}
\put(38,13){$\bullet$}
\put(34,21){${\Gamma_{\hspace{-.08cm}A}}_{P_4}$}
\put(30,10){\vector(0,-1){2}}
\end{picture}}
}
\end{picture}
=
{1 \over 2}
\left[
%
\begin{picture}(55,25)(0,0)
\put(0,-15){
\put(20,10){\begin{picture}(40,20)(0,0)
\put(0,15){\line(1,0){20}}
\put(0,5){\vector(1,0){10}}
\put(25,15){\vector(-1,0){5}}
\put(15,5){\line(-1,0){5}}
\end{picture}}
\put(-5,10){\begin{picture}(40,20)(0,0)
\put(40,10){\oval(30,10)[r]}
\put(42,2){$\bullet$}
\put(38,-5){${\Gamma_{\hspace{-.08cm}A}}_{P_2}$}
\end{picture}}
\put(-25,10){\begin{picture}(40,20)(0,0)
\put(45,10){\oval(30,10)[l]}
\put(38,2){$\bullet$}
\put(30,-5){${\Gamma_{\hspace{-.08cm}A}}_{P_3}$}
\put(45,13){$\bullet$}
\put(30,22){$({\Gamma_{\hspace{-.08cm}A}}_{P_4} {\gamma_5}_{P_1})$}
\put(30,10){\vector(0,-1){2}}
\end{picture}}
}
\end{picture}
+
%
\begin{picture}(55,20)(0,0)
\put(0,-15){
\put(20,10){\begin{picture}(40,20)(0,0)
\put(0,15){\line(1,0){5}}
\put(20,5){\vector(1,0){5}}
\put(15,15){\vector(-1,0){10}}
\put(25,5){\line(-1,0){25}}
\end{picture}}
\put(-5,10){\begin{picture}(40,20)(0,0)
\put(40,10){\oval(30,10)[r]}
\put(42,13){$\bullet$}
\put(38,21){${\Gamma_{\hspace{-.08cm}A}}_{P_1}$}
\end{picture}}
\put(-25,10){\begin{picture}(40,20)(0,0)
\put(45,10){\oval(30,10)[l]}
\put(45,2){$\bullet$}
\put(30,-5){$({\gamma_5}_{P_2}{\Gamma_{\hspace{-.08cm}A}}_{P_3})$}
\put(38,13){$\bullet$}
\put(30,21){${\Gamma_{\hspace{-.08cm}A}}_{P_4}$}
\put(30,10){\vector(0,-1){2}}
\end{picture}}
}
\end{picture}
\right]
\nonumber \\
&&+
{1 \over 4} \bigl[ \ \
\begin{picture}(160,15)(0,0)
\put(0,-18){
\put(105,10){\begin{picture}(40,20)(0,0)
\put(40,10){\oval(12,12)}
\put(46,10){\vector(0,1){2}}
\put(32,8){$\bullet$}
\put(-102,10){$ 
{\gamma_5}_{P_3}{\gamma_5}_{P_4} 
({\gamma_5}_{P_1}{\gamma_{\hspace{-.08cm}A}}_{P_2} +
{\gamma_{\hspace{-.08cm}A}}_{P_1} {\gamma_5}_{P_2} )
$}
\end{picture}}
}
\end{picture}
\nonumber \\
\label{box with ladder}
&& \ \ \ \ + 2
\begin{picture}(160,15)(0,0)
\put(0,-18){
\put(105,10){\begin{picture}(40,20)(0,0)
\put(40,10){\oval(12,12)}
\put(46,10){\vector(0,1){2}}
\put(32,8){$\bullet$}
\put(-102,10){$ 
 {\gamma_5}_{P_4}
({\gamma_5}_{P_1}{\gamma_{\hspace{-.08cm}A}}_{P_2} +
{\gamma_{\hspace{-.08cm}A}}_{P_1} {\gamma_5}_{P_2} )
{\gamma_5}_{P_3}
$}
\end{picture}}
}
\end{picture}
\\
&& \ \ \ \ \ \, +
\begin{picture}(160,15)(0,0)
\put(0,-18){
\put(105,10){\begin{picture}(40,20)(0,0)
\put(40,10){\oval(12,12)}
\put(46,10){\vector(0,1){2}}
\put(32,8){$\bullet$}
\put(-102,10){$ 
({\gamma_5}_{P_1}{\gamma_{\hspace{-.08cm}A}}_{P_2} +
{\gamma_{\hspace{-.08cm}A}}_{P_1} {\gamma_5}_{P_2} )
{\gamma_5}_{P_3} {\gamma_5}_{P_4}
$}
\end{picture}}
}
\end{picture}
\bigr]
\nonumber \\
&&+
{1 \over 4} 
\begin{picture}(220,15)(0,0)
\put(0,-18){
\put(85,10){
\begin{picture}(40,20)(0,0)
\put(0,15){\line(1,0){5}}
\put(0,5){\vector(1,0){10}}
\put(15,15){\vector(-1,0){10}}
\put(15,5){\line(-1,0){5}}
\put(15,0){\framebox(10,20){}}
\put(25,15){\line(1,0){5}}
\put(25,5){\vector(1,0){10}}
\put(40,15){\vector(-1,0){10}}
\put(40,5){\line(-1,0){5}}
\end{picture}}
%
\put(85,10){
\begin{picture}(40,20)(0,0)
\put(40,10){\oval(10,10)[r]}
\put(43,8){$\bullet$}
\put(49,12){$_{( 
{\gamma_{\hspace{-.08cm}A}}_{P_1}  {\gamma_5}_{P_2}
+ {\gamma_5}_{P_1} {\gamma_{\hspace{-.08cm}A}}_{P_2}  
)}$}
\end{picture}}
\put(40,10){
\begin{picture}(40,20)(0,0)
\put(45,10){\oval(10,10)[l]}
\put(38,8){$\bullet$}
\put(-42,12){$_{( 
{\gamma_{\hspace{-.08cm}A}}_{P_3}  {\gamma_5}_{P_4}
+ {\gamma_5}_{P_3} {\gamma_{\hspace{-.08cm}A}}_{P_4}  
)}$}
\end{picture}}
}
\end{picture}
\ ,
\nonumber
\end{eqnarray}
where there are three classes of terms, respectively with
zero, one and two vertices $\gamma_{\hspace{-.08cm}A}$.
I note that all the other factors 
( $\Gamma_{\hspace{-.08cm}A}$, $S$, and the scalar and vector ladder) 
are finite and carry the scale of the effective quark-quark
interaction. The fist term, with zero $\gamma_{\hspace{-.08cm}A}$,
cancels when the three diagrams of eq. (\ref{ciclic 1}) are summed.
Using eq. (\ref{bare}), the second term of 
eq. (\ref{box with ladder}),
with one $\gamma_{\hspace{-.08cm}A}$, is,
\begin{eqnarray}
&& 
{1 \over 4 \, i} \, tr \Bigl\{ 
(\not P_1 - \not P_2) \left[ S_{P_2,P_3} -2  S_{P_3,P_4}+S_{P_4,P_1} \right] 
\nonumber \\
&& \ \ \ \ \
-4 m\left[ S_{P_2,P_3} +2  S_{P_3,P_4}+S_{P_4,P_1} \right] 
\Bigr\} 
\ ,
\label{propamoma}
\end{eqnarray}
where the quark propagators $S_{P_i,P_j}$ are indexed with the attached external 
momenta for book keeping. Expanding up to second order in 
$P_i$ and $M_\pi$, and using the Gell-Mann Oakes and Renner relation (\ref{GMOR}),
the eq. (\ref{propamoma}) simplifies to,
\begin{equation}
{1 \over 4 \, i} \, tr \Bigl\{ (\not P_1 - \not P_2) (P_3-P_4)^\mu \partial_\mu S
\Bigr\} - 4 \, i \, f_\pi^2 \, M_\pi^2 \ .
\label{quasi gagagaga}
\end{equation}
The third term with of eq. (\ref{box with ladder}), 
with two $\gamma_{\hspace{-.08cm}A}$ vertices, is equal to,
\begin{equation}
{ 1 \over 4} 
\begin{picture}(120,20)(0,3)
\put(0,0){
\put(35,0){
\begin{picture}(40,20)(0,0)
\put(0,15){\line(1,0){5}}
\put(0,5){\vector(1,0){10}}
\put(15,15){\vector(-1,0){10}}
\put(15,5){\line(-1,0){5}}
\put(15,0){\framebox(10,20){}}
\put(25,15){\line(1,0){5}}
\put(25,5){\vector(1,0){10}}
\put(40,15){\vector(-1,0){10}}
\put(40,5){\line(-1,0){5}}
\end{picture}}
%
\put(35,0){
\begin{picture}(40,20)(0,0)
\put(40,10){\oval(10,10)[r]}
\put(43,8){$\bullet$}
\put(49,5){$\not P_1- \not P_2 \over i$}
\end{picture}}
%
\put(-10,0){
\begin{picture}(40,20)(0,0)
\put(45,10){\oval(10,10)[l]}
\put(38,8){$\bullet$}
\put(8,5){$ \not P_3- \not P_4 \over i$}
\end{picture}}
}
\end{picture}
\ .
\label{third term}
\end{equation}
It is interesting to remark that the vector Ward identity cancels up to second order 
the momentum dependent part of eq. (\ref{quasi gagagaga})  with 
eq. (\ref{third term}).
The vector Ward produces equations comparable with eq. (\ref{axialWI}) and 
eq. (\ref{second}), in particular,
\begin{equation}
\begin{picture}(83,30)(0,0)
\put(0,0){
\put(13,0){
\begin{picture}(40,20)(0,0)
\put(0,15){\line(1,0){5}}
\put(0,5){\vector(1,0){10}}
\put(15,15){\vector(-1,0){10}}
\put(15,5){\line(-1,0){5}}
\put(15,0){\framebox(10,20){}}
\put(25,15){\line(1,0){5}}
\put(25,5){\vector(1,0){10}}
\put(40,15){\vector(-1,0){10}}
\put(40,5){\line(-1,0){5}}
\end{picture}}
%
%
\put(-2,0){
\begin{picture}(40,20)(0,0)
\put(15,10){\oval(10,10)[l]}
\put(08,8){$\bullet$}
\put(-2,5){$ \not P \over i$}
\end{picture}}
%
\put(10,0){
\begin{picture}(40,20)(0,0)
\put(33,20){$k+P/2$}
\put(33,-8){$k-P/2$}
\end{picture}}
}
\end{picture}
=
S(k+P/2)-S(k-P/2) \ .
\label{vector WI not expanded}
\end{equation}
The expansion up to first order in $P^\mu$ of eq. (\ref{vector WI not expanded})
produces,
\begin{equation}
\begin{picture}(150,20)(0,3)
\put(0,0){
\put(100,0){\begin{picture}(40,20)(0,0)
\put(0,15){\line(1,0){5}}
\put(0,5){\vector(1,0){10}}
\put(15,15){\vector(-1,0){10}}
\put(15,5){\line(-1,0){5}}
\put(15,0){\framebox(10,20){}}
\put(25,15){\line(1,0){5}}
\put(25,5){\vector(1,0){10}}
\put(40,15){\vector(-1,0){10}}
\put(40,5){\line(-1,0){5}}
\end{picture}}
%
\put(53,0){\begin{picture}(40,20)(0,0)
\put(0,15){\line(1,0){5}}
\put(0,5){\vector(1,0){10}}
\put(15,15){\vector(-1,0){10}}
\put(15,5){\line(-1,0){5}}
\put(0,10){\oval(10,10)[l]}
\put(-8,8){$\bullet$}
\put(-45,10){$-i \partial_\mu S^{-1} $}
\end{picture}}
%
\put(55,0){\begin{picture}(40,20)(0,0)
\put(45,10){\oval(10,10)[l]}
\put(37,8){$\bullet$}
\put(15,10){$+ \,\gamma_\mu$}
\end{picture}}
}
\end{picture}
=0 \ .
\label{expanded vector WI}
\end{equation} 
Therefore the sum of the momentum dependent parts of eq. (\ref{quasi gagagaga})  and  
eq. (\ref{third term}),
\begin{equation}
%
\begin{picture}(217,35)(0,8)
\put(0,0){
\put(130,10){\begin{picture}(40,20)(0,0)
\put(0,15){\line(1,0){5}}
\put(0,5){\vector(1,0){10}}
\put(15,15){\vector(-1,0){10}}
\put(15,5){\line(-1,0){5}}
\put(15,0){\framebox(10,20){}}
\put(25,15){\line(1,0){5}}
\put(25,5){\vector(1,0){10}}
\put(40,15){\vector(-1,0){10}}
\put(40,5){\line(-1,0){5}}
\end{picture}}
%
\put(83,10){\begin{picture}(40,20)(0,0)
\put(0,15){\line(1,0){5}}
\put(0,5){\vector(1,0){10}}
\put(15,15){\vector(-1,0){10}}
\put(15,5){\line(-1,0){5}}
\put(0,10){\oval(10,10)[l]}
\put(-8,8){$\bullet$}
\put(-45,10){$-i \partial_\mu S^{-1} $}
\end{picture}}
%
\put(85,10){\begin{picture}(40,20)(0,0)
\put(45,10){\oval(10,10)[l]}
\put(37,8){$\bullet$}
\put(15,10){$+ \,\gamma_\mu$}
\end{picture}}
%
\put(0,10){\begin{picture}(40,20)(0,0)
\put(0,10){${P_3^\mu-P_4^\mu \over i}$}
\put(32,10){$\Biggl[$}
\end{picture}}
%
\put(170,10){\begin{picture}(40,20)(0,0)
\put(10,10){\oval(10,10)[r]}
\put(13,8){$\bullet$}
\put(19,10){${\not P_1 - \not P_2 \over i} $}
\put(3,10){$\Biggr]$}
\end{picture}}
}
\end{picture}
\, ,
\end{equation} 
cancels due to the vector Ward Identity (\ref{expanded vector WI}).
Only the constant term remains, and this is exact up to order $P_i P_j$.
The final result for the Feynman diagrams of eq. (\ref{ciclic 1}) is,
\begin{equation}
- 8 \, i \, n_\pi^2 \, M_\pi^2 \ .
\label{last ciclic 1}
\end{equation}

%
\par
\subsection{ Scattering parameters }

Summing the contributions of eqs. (\ref{last ciclic 1}), (\ref{last ciclic 2}),
(\ref{last ciclic 3}) and (\ref{last ciclic 4}), the Feynman loop of eq. (\ref{loop}) 
results in,
\begin{eqnarray}
&
+3 \left({ 1 \over 2 i\, n_\pi}\right)^4 
( -8 \, i \, n_\pi^2 M_\pi^2)
&
\nonumber \\
&
-2 \left({ 1 \over 2 i\, n_\pi}\right)^3 \sum_{ 4 \, perm.} 
\, 2 \, n_\pi ( P_1^2 + P_1 \cdot P_3 -2 \, M_\pi^2 )
&
\nonumber \\
&
+ \left({ 1 \over 2 i\, n_\pi}\right)^2 \sum_{ 4 \, perm.}
{P_1^2 + P_2^2 +P_1 \cdot P_2 
+P_1 \cdot P_3 +P_2 \cdot P_4 -2 \, M_\pi^2 \over i }
&
\nonumber \\
&
+ \left({1 \over 2 i\, n_\pi}\right)^2 \sum_{ 2 \, perm.}
{P_1^2 + P_3^2 -2 \, M_\pi^2 \over i }
&
\nonumber \\
&
={i \over 2 f_\pi^2} \left[
(P_1+P_2)^2+(P_1+P_4)^2
 -2 \, M_\pi^2 \right]
\ ,
&
\label{loop result}
\end{eqnarray}
where the conservation
$P_1+P_2+P_3+P_4=0$ of momentum 
was used to simplify the result.

\par
\begin{figure}
\begin{picture}(220,40)(0,0)
\put(0,0){
%
\put(100,20){
\begin{picture}(40,20)(0,0)
\put(10,11){\line(1,1){12}}
\put(10,9){\line(1,1){12}}
\put(10,11){\line(1,-1){12}}
\put(10,9){\line(1,-1){12}}
\put(10,11){\line(-1,1){12}}
\put(10,9){\line(-1,1){12}}
\put(10,11){\line(-1,-1){12}}
\put(10,9){\line(-1,-1){12}}
\end{picture}}
\put(100,20){
\begin{picture}(40,20)(0,0)
\put(10,10){\circle*{10}}
\put(-30,20){$P_4 \rightarrow$}
\put(28,20){$\leftarrow P_1$}
\put(28,-6){$\leftarrow P_2$}
\put(-30,-6){$P_3 \rightarrow$}
\end{picture}}
\put(0,20){
\begin{picture}(40,20)(0,0)
\put(0,20){$\leftarrow \, q_{o1},\tau_{o1}$}
\put(7,5){OUT}
\put(0,-6){$\leftarrow \, q_{o2},\tau_{o2}$}
\end{picture}}
\put(170,20){
\begin{picture}(40,20)(0,0)
\put(0,20){$q_{i1},\tau_{i1} \, \leftarrow $}
\put(27,5){IN}
\put(0,-6){$q_{i2},\tau_{i2} \, \leftarrow $}
\end{picture}}
}
\end{picture}
\label{6 combinations}
\caption{ To compute the $T$ matrix for $\pi-\pi$ scattering,
the external momenta of the Feynman loop (\ref{loop})
must be matched with the incoming and outgoing 
pion momenta. There are six different possible combinations.
The  Feynman loop (\ref{loop})
is represented by the full circle. } 
\end{figure}
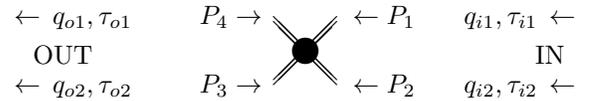

I finally compute the $\pi-\pi$ scattering matrix T.
The external pions, $i1$ and $i2$ 
incoming and $o1$ and $o2$
outgoing, are simply matched
with the four pion vertex that
I just computed in eq.(\ref{loop result}).
This is depicted in Fig. \ref{6 combinations},
where the loop (\ref{loop}) is represented by 
the full circle. The loop is 
topologically invariant for cyclic permutations of 
$P_1,P_2,P_3$ and $P_4$. 
To remove double counting one match
is fixed, say  $P_1=q_{i1}$. Then there are
six different combinations of the remaining
external legs,
\begin{eqnarray}
neighbor :&&
\nonumber \\
P_1=q_{i1} \, , & \ P_2=q_{i2} \, , \ P_3=-q_{o2} & \, , \ P_4=-q_{o1} \ ;
\nonumber \\
P_1=q_{i1} \, , & \ P_2=q_{i2} \, , \ P_3=-q_{o1} & \, , \ P_4=-q_{o2} \ ;
\nonumber \\
P_1=q_{i1} \, , & \ P_2=-q_{o1} \, , \ P_3=-q_{o2} & \, , \ P_4=q_{i2} \ ;
\nonumber \\
P_1=q_{i1} \, , & \ P_2=-q_{o1} \, , \ P_3=q_{i2} & \, , \ P_4=-q_{o2} \ ;
\nonumber \\
separated :&&
\nonumber \\
P_1=q_{i1} \, , & \ P_2=-q_{o2} \, , \ P_3=q_{i2} & \, , \ P_4=-q_{o1} \ ;
\nonumber \\
P_1=q_{i1} \, , & \ P_2=-q_{o2} \, , \ P_3=-q_{o1} & \, , \ P_4=q_{i2} \ ;
\label{all combinations}
\end{eqnarray}
where in the first four combinations the pair of incoming pions
are {\em neighbor } in the Feynman loop, while in
the last two combinations the pair of incoming pions are {\em separated} in the
quark loop by an outgoing pion. 

\par
In what concerns color, all the combinations are
identical because the pion is a color singlet, and the color
factor is appropriately included in the definition of 
$f_\pi$, see eq. (\ref{decay}). 
In what concerns momentum, the result is expressed 
in the usual Mandelstam
relativistic invariant variables $s, t$ and $u$.
For instance the first combination in eq. (\ref{all combinations})
produces the result $ i( s + u -2 M_\pi^2)/( 4 f_\pi^2 )$.
I now introduce flavor. For compactness it was not regarded
in the previous definitions of the vertices ${\Gamma_A}_P$
and $\chi_{_P}$. Because the pion is an isovector,
There are three different cases $I=0$, $I=1$ and $I=2$. 
The flavor contributions to the pion vertex
simply factorize from the momentum contribution,
and the different combinations only produce two classes
of flavor traces, which correspond to the {\em neighbor } 
and {\em separated} classes in eq. (\ref{all combinations}).
The flavor results are compiled in Table \ref{flavor}.
Summing the six possible combinations of color, spin, momentum 
and flavor traces, 
and dividing by $-i$, the $\pi-\pi$
scattering $T^I$ matrices are finally,
\begin{eqnarray}
T^0=& 
- { 2s- M_\pi^2\over 2 f_\pi^2}& - {s+t+u-4 M_\pi^2 \over 2 f_\pi^2} \ ,
\nonumber \\
T^1=& 
- { t-u \over 2 f_\pi^2}& \ ,
\nonumber \\
T^2=&
- { -s+2M_\pi^2\over 2 f_\pi^2}& \, - {s+t+u-4 M_\pi^2 \over 2 f_\pi^2 } \ ,
\label{T matrix}
\end{eqnarray}
where $s+t+u-4 M_\pi^2$ expresses the off mass shell contribution.
The $T^I$ matrices of eq. (\ref{T matrix}) are computed at the tree level 
(including scalar and vector s, t, and u exchange), which 
is exact up to the order of $P_i^2 P_j^2$ and of $M_\pi^2$.
Eq. (\ref{T matrix}) complies with the Gasser and Leutwyler results.
\cite{Leutwyler}.
Off mass shell effects are very important for the
experiments. For instance in the scattering at $\pi-\pi$ 
threshold of a $\pi$ beam with virtual $\pi^*$
provided by a nucleon target 
\cite{Gutay}
, the off mass 
shell effects of eq. (\ref{T matrix}) decrease $T^0$ by a 
factor of 0.5 and increase $T^2$ by a factor of 1.7.

\par
The $\pi-\pi$ scattering lengths $a_0^I$ 
are simply obtained from the mass shell scattering amplitudes 
$T^0,T^2$ with the Born factor of $-1 \over 16 \pi M_\pi$,
and for vanishing 3-momenta. 
The $I=1$ case is antisymmetric so the first
scattering parameter is $a_1^1$
and the corresponding factor is 
${-1 \over 16 \pi M_\pi} { 4 \over 3 (t-u)}$, 
\begin{eqnarray}
a^0_0&=& {7 \over 32 \pi}{ M_\pi \over f_\pi^2} \ ,
\nonumber \\
a^1_1&=& {1 \over 24 \pi}{ 1 \over M_\pi \, f_\pi^2} \ ,
\nonumber \\
a^2_0&=&{-1 \over 16 \pi}{ M_\pi \over f_\pi^2} \ ,
\end{eqnarray}
this is the result of the famous Weinberg theorem for 
$\pi-\pi$ scattering
\cite{Weinberg}.

\par
For simplicity, renormalization 
\cite{Maris,Szczepaniak} 
was omitted from the details of this paper,
and all the Feynman loops were assumed to be finite.
Nevertheless the results are not affected by the use of an ultraviolet divergent kernel, 
say by the one gluon exchange interaction $K(q) \ \alpha \ 1/q^2$ which renders the 
integral in eq. (\ref{mass gap}) logarithmicly divergent. 
The same renormalization of ultraviolet divergences also occurs in atomic 
physics, where the spectrum and the cross section of atoms are simply 
not affected by the divergences which are present in the Shwinger and Dyson 
equation of QED.
Following reference \cite{Maris}, the divergent integral is regularized with an 
ultraviolet cutoff $\Lambda_{UV}$, and the dressed propagator is renormalized in eq. 
(\ref{mass gap}) by the quark wave function renormalization 
factor of $Z_2$ and by the vertex renormalization factor of $Z_1$. 
The Axial Ward Identity (\ref{axialWI}) , 
the Vector Ward Identity  (\ref{expanded vector WI}) ,
and the normalization condition of the Bethe Salpeter vertex (\ref{normalizing}),
ensure that both the axial vertex $\Gamma_{\hspace{-.08cm}A}$,
the vector vertex $V$, 
and the normalization condition of the Bethe Salpeter vertex $\chi_\pi$ 
get the same renormalization factor, which coincides
with the inverse of the renormalization factor of the quark propagator. 
Thus the Adler Zero and the Weinberg result, which are computed in
Feynman loops with an identical number of quark propagators
and vertices (axial, vector or Bethe Salpeter) are insensitive
to the renormalization factors. The results maintain the same
expressions in terms of the physically observable $M_\pi$ and $f_\pi$, which 
are not affected by the ultraviolet infinities in the renormalization factors.

\section{Conclusion} 

\par
In this paper a low momentum and chiral expansion is 
performed on amplitudes computed in the framework of the Quark 
Model with chiral symmetry breaking. The expansion is
analytical, and dressed Feynman diagrams are used for the 
compactness of the expansion. The axial Ward identities,
including a non-trivial Ward identity for the ladder, and 
the vector Ward identity are essential tools to arrive at 
simple and model independent results.

\par
A detailed proof is show, which confirms that 
Goldstone bosons non only are massless, in agreement with
the Gell-Mann, Oakes and Renner relation, but also
that Goldstone bosons are noninteracting at low energy
in agreement with the Adler self-consistency zeroes. 
This proof also confirms that intermediate ladders, 
which describe both meson exchanges and contact terms, 
must be included in low energy quark loops. 
 
\par
Moreover this paper verifies that the famous PCAC relations of 
Goldberger and Treiman and of Weinberg are also obtained 
when the pions are off the mass shell and have a finite size. 
The Quark Model provides a well defined prescription for the 
exchange of virtual intermediate off mass shell pions.

\par
The most involved technical part of this paper consists in
detailing and completing the proof of 
the Weinberg Theorem 
\cite{Weinberg}
which was recently published 
\cite{pi-pi,Goncalo,Steve}, 
and in extending the proof to off mass shell pions.
Other important precursors of this work are the study of
$\pi-\pi$ scattering 
\cite{Veronique} 
in the Nambu and Jona-Lasinio Model,
and the study of $\pi-\pi$ scattering
with the bosonization method
\cite{Roberts}.  

\par
It is a remarkable achievement of chiral symmetry that 
the quark propagator, the geometrical
series of the ladder and the pion Bethe-Salpeter vertex, 
which are functions of the finite scale of the interaction, say the 
string tension $\sigma$ or $\Lambda_{QCD}$,
of the current quark mass $m$, 
and of the ultraviolet cutoff $\Lambda_{UV}$,
explicitly disappear from the final results which 
are simple functions of $f_\pi$ and $M_\pi$ only. 
Any Quark Model with a chirally symmetric
interaction complies with the PCAC relations.
This result is general and model independent. 

\par
I expect that the analytical techniques used here
may also be applied to address other chiral
effects within the quark model framework. 
The determination of the next order terms in the pion momenta 
expansion of the $\pi-\pi$ scattering amplitude will
compute the $l_1$ and $l_2$ parameters of Chiral 
Lagrangians, and will also tests Quark Models \cite{Felipe}. 
Another different extension of this paper would consist in 
addressing the anomalous axial Ward Identities. 
The bosonization method suggests  
\cite{Roberts4}
that the Wess-Zumino term of the pion effective Lagrangean
\cite{Wess,Witten},  
which includes the Levi-Civita symbol $\epsilon^{\mu\nu\alpha\beta}$,
say in the coupling of five pions
\cite{Roberts4},
and the anomalous coupling of pions to photons 
\cite{Schechter,Bando,Roberts5}
can also be studied within this framework. 

\acknowledgements

\par
I mainly acknowledge Emilio Ribeiro for reporting 
on non-trivial Ward identities.
I also acknowledge discussions with Steve Cotanch, Gast\~ao Krein,
Felipe Llanes, Brigitte Hiller, Pieter Maris, Gon\c{c}alo Marques, 
Emilio Ribeiro, and Adam Szczepaniak. 

\par

\pagebreak


%
\begin{table}[t]
\caption{
Table of the flavor traces. 
$tr\{ \tau_{i1} \tau_{i2}\tau_{o2}^\dagger\tau_{o1}^\dagger\}$
and
$tr\{  \tau_{i1}\tau_{o2}^\dagger\tau_{i2}\tau_{o1}^\dagger\}$
are examples of the neighbor and separated cases.
}
\label{flavor}
\begin{tabular}{cccc}
$I \, m_I$ & $\tau_{i1} \tau_{i2}$ & neighbor & separated
\\
\tableline
$0\,0$ & $ { \vec \sigma \cdot \vec \sigma \over 2 \sqrt{6} } $ 
& ${3 \over 4}$ & $-{1 \over 4}$ 
\\
$1\,1$ & $ { \sigma_1 \sigma_2 - \sigma_2 \sigma_1 \over 4 } $ 
& ${1 \over 2}$ & $0$
\\
$2\,2$& $ { \sigma^+ \, \sigma^+ \over \sqrt{2}}$ 
& $0$ & ${1 \over 2}$ 
\end{tabular}
\end{table}
%


%
\begin{figure}
\label{offshell}
\begin{picture}(230,80)(0,0)
\put(0,0){
%
\put(20,50){
\begin{picture}(40,20)(0,0)
\put(10,10.8){\line(2,1){30}}
\put(10,9.2){\line(2,1){30}}
\put(10,11){\line(1,-1){15}}
\put(10,9){\line(1,-1){15}}
\put(10,11){\line(-1,1){15}}
\put(10,9){\line(-1,1){15}}
\put(10,11){\line(-1,-1){15}}
\put(10,9){\line(-1,-1){15}}
\put(10,10){\circle*{10}}
\put(-18,20){$\pi$}
\put(48,20){$\pi$}
\put(23,4){$\pi^*$}
\put(-18,-6){$\pi$}
\end{picture}}
\put(40,30){
\begin{picture}(40,20)(0,0)
\put(10,12){\line(1,-1){15}}
\put(10,10){\line(1,-1){15}}
\put(10,8){\line(1,-1){15}}
\put(10,11){\line(-1,1){15}}
\put(10,9){\line(-1,1){15}}
\put(10,11.6){\line(-2,-1){30}}
\put(10,10){\line(-2,-1){30}}
\put(10,8.4){\line(-2,-1){30}}
\put(10,10){\circle*{10}}
\put(28,-6){$n$}
\put(-38,-6){$n$}
\end{picture}}
%
%
\put(140,40){
\begin{picture}(40,20)(0,0)
\put(10,10.65){\line(1,0){50}}
\put(10,9.35){\line(1,0){50}}
\put(10,10.65){\line(-1,0){20}}
\put(10,9.35){\line(-1,0){20}}
\put(10,11.4){\line(-2,3){15}}
\put(10,8.6){\line(-2,3){15}}
\put(10,11.4){\line(-2,-3){15}}
\put(10,8.6){\line(-2,-3){15}}
\put(10,10){\circle*{10}}
\put(40,10){\circle*{10}}
\put(-18,30){$\pi$}
\put(-23,7){$\pi$}
\put(-18,-16){$\pi$}
\put(58,20){$K$}
\put(23,20){$\pi^*$}
\end{picture}}
\put(40,10){(a)}
\put(160,10){(b)}
}
\end{picture}
\caption{In (a) a $\pi$ on mass shell is scattered by a virtual $\pi^*$ provided
by a nucleon. In (b) a virtual pion, which results from a weak flavor change in
the incoming $K$, decays into three pions. Both (a) and (b) contribute to hadronic 
reactions which are measured in the laboratory.  }
\end{figure}
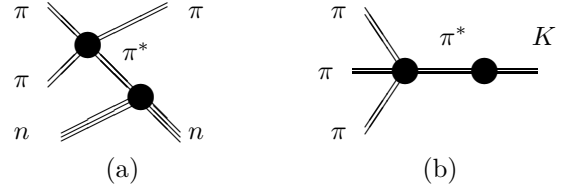

\begin{figure}
\label{Goldberger Treiman}
\begin{picture}(230,90)(0,0)
\put(0,0){
%
%
\put(25,40){
\begin{picture}(40,20)(0,0)
\put(10,10){\line(1,0){30}}
\put(10,10){\line(-1,0){15}}
\put(40,10){\vector(-1,0){10}}
\put(10,10){\vector(-1,0){10}}
\put(-13,12){$u$}
\put(-28,0){$P$}
\put(30,15){$d$}
\put(58,0){$N$}
\end{picture}}
\put(25,40){
\begin{picture}(40,20)(0,0)
\put(20,10){\line(-2,1){30}}
\put(-10,25){\vector(2,-1){17}}
\put(20,10){\line(-2,3){18}}
\put(20,10){\vector(-2,3){13}}
\put(-10,35){$e^-$}
\put(-22,25){$\bar \nu_e$}
\end{picture}}
\put(25,30){
\begin{picture}(40,20)(0,0)
\put(10,10){\line(1,0){40}}
\put(10,10){\line(-1,0){25}}
\put(30,10){\vector(-1,0){15}}
\end{picture}}
\put(25,20){
\begin{picture}(40,20)(0,0)
\put(10,10){\line(1,0){30}}
\put(10,10){\line(-1,0){15}}
\put(30,10){\vector(-1,0){15}}
\end{picture}}
\put(25,20){
\begin{picture}(40,20)(0,0)
\put(17.5,20){\oval(65,20)}
\put(-17.5,18){$\bullet$}
\put(47.5,18){$\bullet$}
\end{picture}}
%
%
%
\put(155,60){
\begin{picture}(40,20)(0,0)
\put(20,10){\line(-1,0){28}}
\put(-10,10){\vector(1,0){15}}
\put(20,10){\line(-3,2){20}}
\put(20,10){\vector(-3,2){15}}
\put(-15,20){$e^-$}
\put(-23,10){$\bar \nu_e$}
\end{picture}}
\put(155,60){
\begin{picture}(40,20)(0,0)
\put(20,-5){\oval(10,29)[t]}
\put(10,-15){\framebox(20,10){}}
\put(25,0){\vector(0,1){5}}
\put(15,5){\vector(0,-1){5}}
\end{picture}}
\put(155,30){
\begin{picture}(40,20)(0,0)
\put(40,10){\line(-3,1){15}}
\put(15,15){\line(-3,-1){15}}
\put(40,10){\vector(-3,1){5}}
\put(15,15){\vector(-3,-1){13}}
\put(-5,12){$u$}
\put(-28,0){$P$}
\put(40,12){$d$}
\put(58,0){$N$}
\end{picture}}
\put(155,20){
\begin{picture}(40,20)(0,0)
\put(25,10){\line(1,0){25}}
\put(15,10){\line(-1,0){25}}
\put(15,10){\vector(-1,0){15}}
\put(50,10){\vector(-1,0){15}}
\end{picture}}
\put(155,10){
\begin{picture}(40,20)(0,0)
\put(25,10){\line(1,0){15}}
\put(15,10){\line(-1,0){15}}
\put(15,10){\vector(-1,0){15}}
\put(40,10){\vector(-1,0){5}}
\end{picture}}
\put(155,10){
\begin{picture}(40,20)(0,0)
\put(20,15){\circle{14}}
\end{picture}}
\put(155,10){
\begin{picture}(40,20)(0,0)
\put(0,20){\oval(20,20)[l]}
\put(40,20){\oval(20,20)[r]}
\put(-12.5,18){$\bullet$}
\put(47.5,18){$\bullet$}
\end{picture}}
\put(40,10){(a)}
\put(170,5){(b)}
}
\end{picture}
\caption{
This figure shows the microscopic description of the weak decay 
$ N \rightarrow P+e^-+\bar \nu_e$ where a $d$ quark produces a $u$ quark and two leptons
with the four Fermi weak coupling. The bare diagram is depicted in (a) 
and the corresponding fully dressed diagram is depicted in (b). (b) includes a full ladder, 
which is represented with a full box, and the series of interactions of the 
diquark which is not coupled to the leptons, which is represented with an empty circle.
} 
\end{figure}
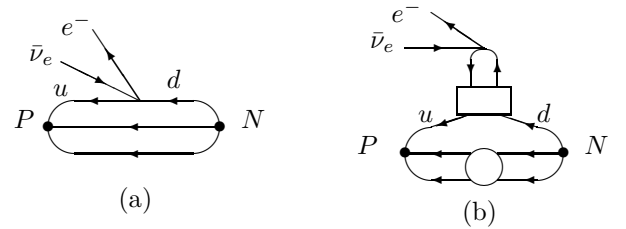

%

\begin{references}
%
\bibitem{Rujula}
A.~De~Rujula, H.~Georgi and S.~L.~Glashow,
Phys.\ Rev.\ D {\bf 12}, 147 (1975).
%
\bibitem{Ribeiro}
J.~Ribeiro,
Z.\ Phys.\ C {\bf 5}, 27 (1980).
%
\bibitem{Gastao}
P.~Bicudo, G.~Krein and J.~Ribeiro,
Phys.\ Rev.\ C {\bf 64}, 025202 (2001)
[arXiv:hep-ph/0105289];
%
\bibitem{pi-pi}
P.~Bicudo, S.~Cotanch, F.~Llanes-Estrada, P.~Maris, E.~Ribeiro and A.~Szczepaniak,
Phys.\ Rev.\ D {\bf 65}, 076008 (2002)
[arXiv:hep-ph/0112015].
%
\bibitem{Goncalo}
P.~Bicudo, M.~Faria, G.~Marques, and J.~Ribeiro
[arXiv:nucl-th/0106071].
%
\bibitem{Pene}
A. Le Yaouanc, L. Oliver, O. Pene and J.-C. Raynal,
Phys.\ Rev.\ D {\bf 29}, 1233 (1984);
Phys.\ Rev.\ D {\bf 31}, 137 (1985).
%
\bibitem{Adler2}
S. Adler and A. Davis,
Nucl.\ Phys.\ B {\bf 244}, 469 (1984).
%
\bibitem{Bicudo}
P.~Bicudo and J.~Ribeiro,
Phys.\ Rev.\ D {\bf 42}, 1611 (1990);
Phys.\ Rev.\ D {\bf 42}, 1625 (1990);
Phys.\ Rev.\ D {\bf 42}, 1635 (1990).
%
\bibitem{Gell-Mann}
M.~Gell-Mann, R.~J.~Oakes and B.~Renner,
Phys.\ Rev.\  {\bf 175}, 2195 (1968).
%
\bibitem{Goldberger}
M.~L.~Goldberger and S.~B.~Treiman,
Phys.\ Rev.\  {\bf 111}, 354 (1958).
%
\bibitem{Adler}
S.~L.~Adler,
Phys.\ Rev.\  {\bf 137}, B1022 (1965).
%
\bibitem{Weinberg}
S. Weinberg,
Phys.\ Rev.\ Lett.\  {\bf 17}, 616 (1966).
%
\bibitem{pion size}
S.~R.~Amendolia {\it et al.},
Phys.\ Lett.\ B {\bf 146}, 116 (1984),
E.~B.~Dally {\it et al.},
Phys.\ Rev.\ Lett.\  {\bf 48}, 375 (1982).
%
\bibitem{Colangelo}
G.~ Colangelo,
talk presented at the conConfinement and the Hadron spectrum V,
Gargnano, Italy, September 2002
%
\bibitem{Yazov}
V. ~Yazov, 
talk presented at the Confinement and the Hadron spectrum V,
Gargnano, Italy, September 2002
%
\bibitem{Martin}
P.~Estabrooks and A.~D.~Martin,
Nucl.\ Phys.\ B {\bf 79}, 301 (1974).
%
\bibitem{Llewellyn-Smith}
C. Llewellyn-Smith, Nuo.\ Cim.\ {\bf 60} A, 348 (1969). 
%
\bibitem{Nambu}
Y. Nambu and J. Jona-Lasinio,
Phys.\ Rev.\  {\bf 124}, 246 (1961);
Phys.\ Rev.\  {\bf 122}, 345 (1961).
%
\bibitem{Veronique}
V.~Bernard, U.~G.~Meissner, A.~Blin and B.~Hiller,
Phys.\ Lett.\ {\bf B253}, 443 (1991);
V.~Bernard, A.~H.~Blin, B.~Hiller, Y.~P.~Ivanov, A.~A.~Osipov and U.~Meissner,
Annals Phys.\ {\bf 249}, 499 (1996)
[arXiv:hep-ph/9506309].
%
\bibitem{Liu}
Y. Dai, C. Huang and D. Liu,
Phys.\ Rev.\ D {\bf 43}, 1717 (1991).
%
\bibitem{Roberts2}
C.~D.~Roberts and S.~M.~Schmidt,
Prog.\ Part.\ Nucl.\ Phys.\  {\bf 45}, S1 (2000)
[arXiv:nucl-th/0005064].
%
\bibitem{Smekal}
R.~Alkofer and L.~von Smekal,
Phys.\ Rept.\  {\bf 353}, 281 (2001)
[arXiv:hep-ph/0007355].
%
\bibitem{Sauli}
V. Sauli,
[arXiv:hep-ph/0108160].
%
\bibitem{vector axial}
P. Bicudo, D.-S. Liu, J. Ribeiro, J. Villate
Phys.\ Rev.\ D {\bf 47}, 1145 (1993).
%
\bibitem{Pagels}
W. Marciano and H. Pagels,
Phys.\ Rept.\  {\bf 36}, 137 (1978).
%
\bibitem{Scadron}
R.~Delbourgo and M.~D.~Scadron,
J.\ Phys.\ G {\bf 5}, 1621 (1979).
%
\bibitem{scalar}
P. Bicudo,
Phys.\ Rev.\ C {\bf 60}, 035209 (1999)
[arXiv:nucl-th/9802058].
%
\bibitem{Maris}
P. Maris, C. Roberts, and P. Tandy
Phys.\ Lett.\ B {\bf 420}, 267 (1998)
[arXiv:nucl-th/9707003].
%
\bibitem{Ivanov}
M.A.Ivanov, Yu.L.Kalinovsky and C.D.Roberts,
Phys. Rev. D {\bf 60}, 034018 (1999) 
[arXiv:nucl-th/9812063].
\bibitem{Maris2}
P.Maris, in 
{\it Quark Confinement and the Hadron Spectrum IV}, p.\ 163.
World Scientific, ed. W. Lucha, K. Maung Maung (2002) 
[arXiv:nucl-th/0009064].
\bibitem{Roberts3}
M.B.Hecht, C.D.Roberts and S.M.Schmidt, in
{\it Quark Confinement and the Hadron Spectrum IV}, p.\ 27
World Scientific, ed. W. Lucha, K. Maung Maung (2002) 
[arXiv:nucl-th/0010024];
P. Maris, A. Raya, C.D. Roberts and S.M. Schmidt, 
[arXiv:nucl-th/0208071];
\bibitem{Bloch}
J.C.R. Bloch, C.D. Roberts and S.M. Schmidt
Phys. Rev. C {\bf 61} (2000) 065207, [arXiv:nucl-th/9911068].
\bibitem{Maris3}
P.Maris and P.C.Tandy,
Phys. Rev. C {\bf 61}, 045202 (2000), [arXiv:nucl-th/9910033];
Phys. Rev C {\bf 62}, 055204 (2000), [arXiv:nucl-th/0005015];
Phys. Rev. C {\bf 65}, 045211 (2002), [arXiv:nucl-th/0201017].
\bibitem{Ji}
C.-R. Ji and P. Maris,
Phys. Rev. D {\bf 64} 014032 (2001), nucl-th/0102057.
%
\bibitem{Pagels2}
H.~Pagels,
Phys.\ Rev.\  {\bf 179}, 1337 (1969).
%
\bibitem{Ishii}
N.~Ishii,
Nucl.\ Phys.\ {\bf A689}, 793 (2001)
[arXiv:nucl-th/0004063].
%
\bibitem{miracle}
V. De Alfaro, S. Fubini, G. Furlan, C. Rosseti,
``Currents in Hadron Physics'' ,
Amsterdam, North-Holland, (1973).
%
\bibitem{Leutwyler}
J. Gasser and H. Leutwyler,
Annals Phys.\  {\bf 158}, 142 (1984).
%
\bibitem{Roberts}
C. Roberts, R. Cahill, M. Sevior and N. Iannella,
Phys.\ Rev.\ D {\bf 49}, 125 (1994)
[arXiv:hep-ph/9304315].
%
\bibitem{Gutay}
L.~J.~Gutay, F.~T.~Meiere and J.~H.~Scharenguivel,
Phys.\ Rev.\ Lett.\  {\bf 23}, 431 (1969).
%
\bibitem{Szczepaniak}
A. ~Szczepaniak and E. ~Swanson,
Phys.\ Rev.\ D {\bf 55}, 1578 (1997)
[arXiv:hep-ph/9609525];
Phys.\ Rev.\ D {\bf 62}, 094027 (2000)
[arXiv:hep-ph/0005083];
Phys.\ Rev.\ Lett.\  {\bf D87}, 072001 (2001)
[arXiv:hep-ph/0006306].
%
\bibitem{Steve}
S.Cotanch, P.Maris, [arXiv:hep-ph/0210151].
%
\bibitem{Felipe}
F. Llanes-Estrada and P. Bicudo, in {\it Quark Confinement and the Hadron Spectrum V}, 
World Scientific, ed. N. Brambilla and G. Prosperi (2002) [arXiv:hep-ph/0212182].
%
\bibitem{Roberts4}
J. Praschifka, C.D. Roberts and R.T. Cahill, Phys. Rev. D {\bf 36}, 209 (1987);
C.D. Roberts, R.T. Cahill and J. Praschifka, Annals Phys.(NY) {\bf 188}, 20 (1988).
%
\bibitem{Wess}
J.~Wess and B.~Zumino,
Phys.\ Lett.\ B {\bf 37}, 95 (1971).
%
\bibitem{Witten}
E.~Witten,
Nucl.\ Phys.\  B {\bf 223}, 422 (1983).
%
\bibitem{Schechter}
\"O. Kaymakcalan, S. Rajeev and J. Schechter
Phys. Rev. D {\bf 30}, 594 (1984);
H. Gomm, \"O. Kaymakcalan and J. Schechter
Phys. Rev. D {\bf 30}, 2345 (1984);
\"O. Kaymakcalan and J. Schechter
Phys. Rev. D {\bf 31}, 1109 (1985).
%
\bibitem{Bando}
M.Bando, M.Harada and T.Kugo, Prog. Theor. Phys. {\bf 91}, 927 (1994), [arXiv:hep-ph/9312343].
\bibitem{Roberts5}
C.D. Roberts, Nucl. Phys. A {\bf 605}, 475 (1996), [arXiv:hep-ph/9408233];
R. Alkofer and C.D. Roberts, Phys. Lett. B {\bf 369}, 101 (1996), [arXiv:hep-ph/9510284];
P.Maris and C.D.Roberts, Phys. Rev. C {\bf 58}, 3659 (1998), [arXiv:nucl-th/9804062].
%
\end{references}
\end{document}